\documentclass[a4paper]{JHEP3}
\usepackage{amsmath, amssymb, mathtools, stmaryrd}
\usepackage{amsthm}
\usepackage{amsfonts}
\usepackage{epsfig,multicol,float}

\def\ba{\begin{align}}
\def\ea{\end{align}}
\def\be{\begin{equation}}
\def\ee{\end{equation}}
\def\nn{\nonumber}
\def\bea{\begin{eqnarray}}
\def\eea{\end{eqnarray}}

\def\exd{{\rm d}}

\def\pref#1{(\ref{#1})}

\newcommand{\roughly}[1]{\mathrel{\raise.3ex\hbox{$#1$\kern-0.85em
  \lower1ex\hbox{$\sim$}}}}

\newcommand{\gsim}{\roughly>}

\def\Fd{{\widetilde F}}
\def\Gd{{\widetilde G}}

\def\DBI{{\scriptscriptstyle DBI}}
\def\CFT{{\scriptscriptstyle CFT}}

\def\cF{{\cal F}}
\def\cG{{\cal G}}

\def\cO{{\cal O}}
\def\cT{{\cal T}}

\def\ssS{{\scriptscriptstyle S}}
\def\ssT{{\scriptscriptstyle T}}

\makeatletter
\def\x@arrow{\DOTSB\Relbar}
\def\xlongequalsignfill@{\arrowfill@\x@arrow\Relbar\x@arrow}
\newcommand{\xlongequal}[2]{%
	\ext@arrow 0099\xlongequalsignfill@{#1}{#2}}
\makeatother

\title{Finite Size Scaling in Quantum Hallography}

\author{Allan Bayntun${}^1$ and C.P.~Burgess${}^{1,2}$\\

${}^1$Department of Physics \& Astronomy, McMaster University\\ \qquad 1280 Main Street West, Hamilton ON, Canada.\\

${}^2$Perimeter Institute for Theoretical Physics\\
\qquad 31 Caroline Street North, Waterloo ON, Canada.\\}

\abstract{At low temperatures observations of the Hall resistance for Quantum Hall systems at the interface between two Hall plateaux reveal a power-law behaviour, $\exd R_{xy}/\exd B \propto T^{-p}$ (with $p = 0.42 \pm 0.01$); changing at still smaller temperatures, $T < T_s$, to a temperature-independent value. Experiments also show that the transition temperature varies with sample size, $L$, according to $T_s \propto 1/L$. These experiments pose a potential challenge to the holographic AdS/QHE model recently proposed in {\tt arXiv:1008.1917}. This proposal, which was motivated by the natural way AdS/CFT methods capture the emergent duality symmetries exhibited by quantum Hall systems, successfully describes the scaling exponent $p$ by relating it to an infrared dynamical exponent $z$ with $p = 2/z$. For a broad class of models $z$ is robustly shown to be $z = 5$ in the regime relevant to the experiments (though becoming $z = 1$ further in the ultraviolet). By incorporating finite-size effects into these 
models we show that they reproduce a transition to a temperature-independent regime, predicting a transition temperature satisfying $T_s \propto 1/L$ or $\propto 1/L^5$ in two separate regions of parameter space, even though $z = 5$ governs the temperature dependence of the conductivity in both cases. The possibility of a deviation from naive $z=5$ scaling arises because the brane tension introduces a new scale, which alters where the transition between UV and IR scaling occurs, in an $L$-dependent way.  The AdS/CFT calculation indicates the two regimes of temperature scaling are separated by a first-order transition, suggesting new possibilities for testing the picture experimentally. Remarkably, in this interpretation the gravity dual of the transition from temperature scaling to temperature-independent resistance is related to the Chandrashekar transition from a star to a black hole with increasing mass.
}

\begin{document}

\section{Introduction}

In essence the AdS/CFT correspondence asserts that the space of field theories is smaller than had been previously thought: two theories previously believed to be unrelated to one another turn out to be different descriptions of the same physics \cite{AdSCFT}. This equivalence was not understood earlier because it is the strongly coupled limit of one theory in the pair that is equivalent to the weakly coupled limit of the other, and vice versa. What is most remarkable is how dramatically different the two related theories superficially are: a non-gravitational theory in $d$ dimensions is equivalent to a gravitational theory in $d+1$ dimensions. See ref.~\cite{AdSCFTrev1} (or refs.~\cite{AdSCFTrevs}) for an intuitive (or more detailed) review.

This observation has generated considerable recent interest in applying AdS/CFT methods to condensed matter systems (for reviews, see \cite{AdSCMTrevs}). The desired end-game for these studies is to find a gravity dual that correctly predicts the properties of a system of strongly correlated electrons. The hope is that AdS/CFT methods might shed light on dynamics to which current tools have no access by providing a new class of relatively simple physical models of strongly interacting systems. Quantum Hall systems --- for which strongly correlated electrons exhibit a variety of surprising and remarkable properties \cite{GirvinReview} --- provide a promising point of potential contact for AdS/CFT methods \cite{holographicHall}.

For this paper our interest is in a particular AdS/QHE approach, ref.~\cite{Hallography}, which starts from the observation that transitions among quantum Hall plateaux exhibit a number of robust experimental properties \cite{QHEdualityexps, Wei} that have a simple and universal phenomenological interpretation \cite{DualityInterp} in terms of a class of emergent `symmetries' \cite{LCS}. These symmetries act directly on the Ohmic and Hall conductivities: if $\sigma = \sigma_{xy} + i \sigma_{xx}$ is measured in units of $e^2/h$, then a variety of remarkable observations are consistent with (and derivable from) the statement that the flow of $\sigma$ with changing temperature commutes with the action of a discrete duality group: $\sigma \to (a \, \sigma + b)/(c \, \sigma + d)$, with integers $a$, $b$, $c$ and $d$ satisfying $ad - bc = 1$, with $c$ even. Although evidence has been accumulating over many years \cite{PVD} that (2+1)-dimensional conformal systems very often enjoy such symmetries, this property 
turns out to be
particularly manifest within the AdS/CFT framework \cite{AdSCFTduality, Shamit-attractor}.

Of course, given a class of theories that capture these symmetries, the acid test is to find models that also get other experiments right that do not follow immediately from symmetry considerations. An encouraging feature of the proposal of ref.~\cite{Hallography} is that it correctly models a measured scaling exponent whose numerical value is not simply a consequence of the emergent symmetries. Specifically, as described more fully below, between two Hall plateaux the differential Hall resistance at low temperatures behaves as
\be \label{dRxydBvsT}
 \left( \frac{\exd R_{xy}}{\exd B} \right)_{B_c} \propto T^{-p} \,,
\ee
with $p$ measured to be $p = 0.42 \pm 0.01$ \cite{Wei, QHEfinitesizeexp}. In \cite{Hallography} this is predicted to be $p \simeq 2/z$, where $z$ is an infrared dynamical exponent \cite{ScalingRev} whose value evaluates (for a broad class of models -- see below) to $5$.

But the real power of having an explicit model is that it allows different measurements to be related to one another. In particular, because $z$ is a dynamical exponent ({\em i.e.} describes the relative scaling of time and space as one coarse-grains high-frequency modes), a prediction for $z$ can be tested in other ways besides through its implications for $p$, and these must also agree with experiment. And for quantum Hall systems several other measurements appear to indicate $z=1$ \cite{QHEfinitesizeexp, QHEMicrowaveexp}.

Our purpose in this paper is to argue that the experimental evidence for $z = 1$ is consistent with the AdS/QHE framework proposed in \cite{Hallography}, including its successful description of $p$. There are two separate reasons for this, both of which come down to the precise domain of validity of the theory's reproduction of $z = 5$. First, although the theory allows $z = 5$ for the dynamical scaling exponent in the far infrared, it also predicts a crossover to $z = 1$ in the ultraviolet. So experiments that indicate $z = 1$ in the ultraviolet, such as AC conductivity measurements \cite{QHEMicrowaveexp}, do not actually disagree, even at face value, with the $z=5$ temperature scaling found in the deep infrared by \cite{Hallography}.

More problematic are experiments like \cite{QHEfinitesizeexp}, that find evidence for $z = 1$ directly in the same regime where $p$ is measured. As described in more detail below, these measurements find that at small enough temperatures the scaling behaviour, eq.~\pref{dRxydBvsT}, eventually stops, with $\exd R_{xy}/\exd B$ becoming $T$-independent for $T < T_s$. The evidence for $z = 1$ comes because the crossover temperature between the scaling and $T$-independent regimes is observed to vary with system size, $L$, as $T_s \propto 1/L$.

To address these measurements we extend the analysis of \cite{Hallography} to include finite-size effects, in order to see if the AdS/QHE model properly captures the onset of a $T$-independent regime. We find that it does, predicting a transition to a $T$-independent (but $L$-dependent) Hall resistance for $T < T_s$. More remarkably, we find that the transition between $T$-independence and scaling like $T^{-2/5}$ occurs at a transition temperature that can scale either as $T_s \propto 1/L^5$ or $T_s \propto 1/L$ depending on the size of the tension, $\cT$, on the brane that is required on the AdS side to describe the charge carriers.

Ultimately, the possibility for having two kinds of $L$-scaling for $T_s$ without changing the $T$-scaling of the Hall resistance can be traced to the presence of this brane tension, which provides an intrinsic scale to the problem and so causes deviations from naive scaling behaviour. In particular, we find that $\cT$ causes the transition point between the IR and UV scaling regimes to vary in an $L$-dependent (but not $T$-dependent) way. This allows the transition temperature to scale as $T_s \propto 1/L$ or $T_s \propto 1/L^5$ depending on the size of $\cT$, even though $\exd R_{xy}/\exd B \propto T^{-2/5}$ for $T > T_s$ in both cases.

We incorporate finite-size effects by generalizing standard AdS/CFT methods introduced by ref.~\cite{Witten-gauge}, which argues that black holes are not the appropriate solution for describing finite-size systems at sufficiently low temperatures. In the absence of a chemical potential in the CFT (which corresponds on the gravity side to a black hole with no charge) ref.~\cite{Witten-gauge} argues that the better low-temperature solution --- {\em i.e.} the one with lower free energy --- is empty anti-de Sitter space. Entropy density is discontinuous across the transition to the low-temperature phase indicating that the transition is first order.

Our main generalization of this argument is to the case of nonzero chemical potential, for which the preferred low-temperature solution is instead an electrically charged star, rather than empty anti-de Sitter space. (See \cite{Hartnoll:2010gu, Arsiwalla:2010bt} for similar considerations with a chemical potential in the infinite-volume limit.) Remarkably, this links the quantum Hall transition from temperature-scaling to temperature-independence with the Chandrashekar transition from a star to a black hole as its mass is increased. The transition between the black-hole and stellar phases is again first order, suggesting the possibility of there being experimental tests of this picture if the thermal properties of the electron gas can be accessed (certainly a difficult experimental challenge for samples this small).

We organize our presentation as follows. The remainder of this section does two things: first \S\ref{sec:qhallography} summarizes the AdS/QHE proposal of ref.~\cite{Hallography}; and then \S\ref{sec-QHEfinite} outlines the finite-size experimental results of \cite{QHEfinitesizeexp}. \S\ref{sec:finitesizeADSCFT} then describes how to incorporate finite-size effects into an AdS/CFT framework, starting in \S\ref{sec-adsfinite} with a summary of ref.~\cite{Witten-gauge}'s analysis in terms of a Hawking-Page transition. \S\ref{sec-DBI-finite} then describes the extension of this analysis to the AdS/QHE system, for which the main complication is the nonzero chemical potential. For nonzero chemical potential the transition at low temperatures on the gravity side is pictured to be into a phase described by a charged star. \S\ref{sec-TOVequations} develops the gravity description of the star and what its properties imply for the AdS/QHE system. \S\ref{sec-phasediagram} calculates the free energies of the two phases 
and
demonstrates that the new stellar phase is preferred at sufficiently low temperatures, $T < T_s$.  \S\ref{sec-comparison} then shows that the transition is usually related to system size by $T_s \propto 1/L$, and also identifies specific parts of parameter space for which this behaviour could break down. Finally, \S\ref{sec-conclusion} briefly summarizes our results together with potential future directions.

\subsection{The AdS/QHE system}
\label{sec:qhallography}

The action for the AdS/QHE model proposed in ref.~\cite{Hallography} has the following form
\be
 S = S_{\rm grav} + S_{\rm matter} + S_{\rm probe} \,,
\ee
where the equations of motion for $S_{\rm grav} + S_{\rm matter}$ give the black-brane solution that describes the CFT's thermal properties at infinite volume. $S_{\rm probe}$ describes a probe brane whose charge carriers give rise to the Ohmic and Hall conductivities of interest for quantum Hall phenomenology. By assumption $S_{\rm probe}$ only responds to, and does not perturb, the fields sourced by $S_{\rm grav} + S_{\rm matter}$ (we comment below on the necessity for both $S_{\rm matter}$ and $S_{\rm probe}$).

\subsubsection*{The bulk}

With duality in mind the gravitational sector is chosen to be $SL(2,R)$ invariant,\footnote{Strictly speaking, $SL(2,R)$ is a classical symmetry, and only a discrete subgroup like $SL(2,Z)$ or a smaller subgroup is expected to survive quantum effects \cite{Hallography}.}
\be \label{EinsteinDilatonAction}
 S_{\rm grav} = - \int \exd^4 x \, \sqrt{-g} \;
 \left\{ \frac{1}{2\kappa^2} \left[ R - \frac{6}{L^2}
 + \frac\zeta2
 \left( \partial_\mu \tilde\phi \,\partial^\mu \tilde\phi
 + e^{2\tilde\phi} \; \partial_\mu \chi \, \partial^\mu \chi
 \right) \right] \right\} \,,
\ee
with $SL(2,R)$ acting according to
\be \label{SL2Rtau}
 \tau \to \frac{a \, \tau + b}{c \, \tau + d}
 \quad \hbox{and} \quad
 g_{\mu\nu} \to g_{\mu\nu} \,,
\ee
where $a$, $b$, $c$ and $d$ are arbitrary real numbers that satisfy the $SL(2,R)$ condition $ad - bc = 1$ and
\be
 \tau := \chi + i e^{-\tilde \phi} \,.
\ee
We choose $\zeta = 1$, as supersymmetry would require if we were to embed this into a more complete description.

For the matter sector we take a Maxwell field, $B_\mu$, governed by the $SL(2,R)$-invariant Dirac-Born-Infeld (DBI) lagrangian
\begin{eqnarray}  \label{DBI-dilaton-axion}
 S_{\rm matter} &=& - \int \exd^4x \sqrt{-g} \; \cT
 \Bigl( X - 1 \Bigr)
 -\frac14 \, \int \exd^4x \, \sqrt{-g}
 \; \chi F_{\mu\nu} \Fd^{\mu\nu}  \nn\\
 \hbox{with} \quad
 X &:=& \sqrt{1 + \frac{\ell^4}{2} \, e^{-\tilde\phi} F_{\mu\nu} F^{\mu\nu} - \frac{\ell^8}{16} \,  e^{-2\tilde\phi} \left(
 F_{\mu\nu} \Fd^{\mu\nu} \right)^2 } \,,
\end{eqnarray}
where $F := \exd B$ is the usual 2-form Maxwell field strength and\footnote{In our conventions the Levi-Civita symbol, $\epsilon_{\mu\nu\lambda\rho}$ transforms as a tensor rather than a tensor density.} $\widetilde{F}_{\mu\nu} := \frac12 \, \epsilon_{\mu\nu\lambda \rho} F^{\lambda \rho}$.

Notice that the square-root term reduces to the usual Maxwell action (with a non-minimal dilaton coupling) in the limit $\ell \to 0$ with $\cT \ell^4$ fixed. Unlike the Maxwell action, the DBI action has the advantage of being able to handle nonlinear situations where the conductivities themselves depend on the applied potentials. The action of $SL(2,R)$ on the Maxwell field is most simply written as \cite{GiRa}
\be \label{Maxwelltransfn}
 \left( \begin{array}{c} \cG_{\mu\nu} \\
     \cF_{\mu\nu} \\ \end{array} \right)
 \to
 \left( \begin{array}{cc}
     a & b \\
     c & d \\
   \end{array} \right)
 \left( \begin{array}{c} \cG_{\mu\nu} \\
     \cF_{\mu\nu} \\ \end{array} \right) \,,
\ee
where script fields denote the complex quantities
\be
 \cF_{\mu\nu} := F_{\mu\nu} - i \Fd_{\mu\nu}
 \quad \hbox{and} \quad
 \cG_{\mu\nu} := - \Gd_{\mu\nu} - i G_{\mu\nu} \,,
\ee
with
\be
 G^{\mu\nu} := - \frac{2}{\sqrt{-g}} \, \left( \frac{\delta
 S}{\delta F_{\mu\nu}} \right) = \frac{\cT \ell^4}{X} \, \left[ e^{-\tilde\phi} F^{\mu\nu} - \frac{\ell^4}{4} \, e^{-2 \tilde \phi} (F_{\lambda\rho} \widetilde F^{\lambda\rho})
 \Fd^{\mu\nu} \right] + \chi \Fd^{\mu\nu}  \,.
\ee

$SL(2,R)$ invariance of the field equations allows a great simplification when seeking black-brane/black-hole solutions, since they can always be used to set $\chi$ and any magnetic charge, $Q_m$, to zero. It suffices therefore to consider black branes with electric charge, $Q$, (defined more precisely by \eqref{Qdef}, below) in the presence of a nonvanishing dilaton field. Some simplifications are possible even in this case, however, since the $SL(2,R)$ transformation with $b=c=0$, $a=1/d$ preserves the choice $\chi = Q_m = 0$, while transforming
\be
 \tilde\phi\rightarrow \tilde\phi - 2 \log a
 \quad \hbox{and} \quad
 Q\rightarrow a \, Q \,.
\ee
This shows that $Q$ and $\tilde\phi$ should only appear in the invariant combination $Q^2 \, e^{\tilde\phi}$.

The field equations for this action admit charged black-brane solutions \cite{Hallography} that have the following large- and small-$r$ forms\footnote{The near-brane form shown here relies on the choice $\zeta = 1$, made above.}
\bea
 \exd s^2 &\simeq& h_\infty \, \tilde r^2 \, \exd t^2 + \frac{\exd \tilde r^2}{h_\infty \, \tilde r^2}
 + \tilde r^2 \Bigl(
 \exd x^2 + \exd y^2 \Bigr)
 \qquad \hbox{(large $\tilde r$ or UV)} \\
 \exd s^2 &\simeq& h_0 \, r^{10} \, \exd t^2 + \frac{\exd \tilde r^2}{h_0 \, \tilde r^2} + \tilde r^2 \Bigl(
 \exd x^2 + \exd y^2 \Bigr)
 \qquad \;\hbox{(small $\tilde r$ or IR)}\,,
\eea
where $h_0$ and $h_\infty$ are constants. These solutions closely resemble those found for the non-DBI Maxwell-axio-dilaton theory in ref.~\cite{Shamit-attractor}. Notice in particular they are invariant under rescalings $x \to \lambda \, x$ and $y \to \lambda \, y$ provided $\tilde r \to \tilde r/\lambda$ and $t \to \lambda^z \, t$, with $z = 1$ in the UV (large-$\tilde r$) and $z = 5$ in the IR (small-$\tilde r$) limits.

\subsubsection*{The probe}

The probe system couples to the real electromagnetic field, $A_\mu$, for which we again choose precisely the same DBI action as given in eq.~\pref{DBI-dilaton-axion}.
\be
 S_{\rm probe}(A_\mu) = S_{\rm matter} (B_\mu \to A_\mu) \,,
\ee
but with a much smaller tension (to justify the probe approximation).

It might seem redundant to have two almost identical sectors, $S_{\rm matter}$ and $S_{\rm probe}$, and it probably is. Both sectors are needed in \cite{Hallography} in order to achieve a finite DC conductivity for general magnetic fields, with the CFT described by the black brane providing a source of dissipation for the charge carriers described by $S_{\rm probe}$ \cite{Hartnoll:2009ns}. Without the probe sector the remaining black brane does not break translation invariance and so conservation of momentum suppresses the dissipation required for generating a generic DC resistance.

However from the point of view of the long game this likely only reflects how poorly developed is the present state of the holographic art. It would seem more efficient to drop $S_{\rm matter}$ altogether while simultaneously dropping the probe approximation for $S_{\rm probe}$, so that $S_{\rm probe}$ can play $S_{\rm matter}$'s role in shaping the form of the black-hole geometry.  It is likely that $S_{\rm matter}$ could be dropped in this way once an efficient formulation of disorder becomes available for AdS/CFT systems \cite{AdSorder}.

\subsubsection*{Domain of approximation}

Since our analysis is semiclassical, we must stake out its domain of validity. The first approximation required is weak coupling, which amounts to the requirement
\be
 e^{\tilde\phi} \ll 1 \,.
\ee

Next comes the low-energy/small-curvature approximation that allows one to work within a low-energy field theoretic (gravity) description. As discussed in \cite{Hartnoll:2009ns} this requires the curvature radius, $L$, to be much larger than the string length, $\ell/X$, in the presence of the background Maxwell field: $L \gg \ell/X$.

Of particular interest in what follows are situations where the near-horizon geometry is well-described by the near-horizon $z=5$ solution given above, for which the dilaton profile satisfies $e^{\tilde \phi} \propto \tilde r^4$. In this case $X$ is approximately $\tilde r$-independent and the condition $L \gg \ell/X$ simplifies to
\be  \label{supgravcond}
 1 \gg \frac{\ell^2}{X^2 L^2} \simeq \frac{\ell^2}{L^2}
 \left( 1 + \frac{Q^2e^{\tilde\phi}}{\mathcal T^2 \ell^4
 \tilde r^4} \right) \simeq
 \frac{Q^2e^{\tilde\phi}}{\mathcal T^2 L^2\ell^2 \tilde r^4} \simeq  \frac{Q^2e^{\tilde\phi_h}}{\mathcal T^2 L^2\ell^2 \tilde r_h^4}  \,,
\ee
where $\tilde r_h$ represents the position of the black brane horizon and $\tilde\phi_h := \tilde \phi(r_h)$. Eq.~\pref{supgravcond} uses the approximate expression for $X$, eq.~\pref{Xh nearhorizon form}, that we find from the near-horizon solutions of later sections, together with the observation that $\frac{Q^2e^{\tilde\phi}}{\mathcal T^2 \ell^4 \tilde r^4}$ is typically large in the asymptotic near-horizon regime of interest.

In principle this represents a lower bound as to how small $L$ can be made, which is of interest in the finite-volume case for which $L$ ends up playing the role of system size. In practice, however, no matter how tiny $L$ becomes eq.~\pref{supgravcond} can always be satisfied by demanding the gauge coupling $g^2 := e^{\tilde\phi_h}/(\cT \ell^4)$ to be sufficiently small. It turns out that the dilaton profile grows monotonically with $\tilde r$, and so $e^{\tilde\phi_h} < e^{\tilde \phi_0}$, where $\tilde \phi_0$ is the asymptotic dilaton value at large $\tilde r$. This ensures that small $e^{\tilde \phi_h}$ can be ensured by choosing $e^{\tilde \phi_0}$ to be sufficiently small.

The probe approximation, which we use when calculating the conductivity (see appendix \ref{sec-condcalc}), imposes additional constraints. Denoting the probe-brane tension by\footnote{For instance, if the matter source is a stack of $N$ identical branes then $\cT = N \cT_p$, for some large $N$.} $\cT_p$, we must demand $\kappa^2 (\cT_p/X) \ll 1/L^2$, where $\cT_p/X$ is the size of the DBI brane stress energy and $1/L^2$ is a typical background curvature. Equivalently, $\kappa^2 L^2 \cT_p \ll X$ or
\be
 \mathcal{\hat T}_p \ll \left[ {1 + \frac{Q^2 e^{\tilde\phi}}{\mathcal T^2 \ell^4 \tilde r^4}} \right]^{-1/2}
 \,,
\ee
where $\hat\cT_p := \kappa^2 L^2 \cT_p$.

\subsection{Experimental finite-size effects}
\label{sec-QHEfinite}

This section briefly summarizes the measurements of ref.~\cite{QHEfinitesizeexp}, mentioned in the introduction, that provide evidence for $z = 1$ through the low-energy temperature dependence of the Hall resistance midway between two plateaux.

\FIGURE[bh]{
 \centering
 \includegraphics[width=110mm]{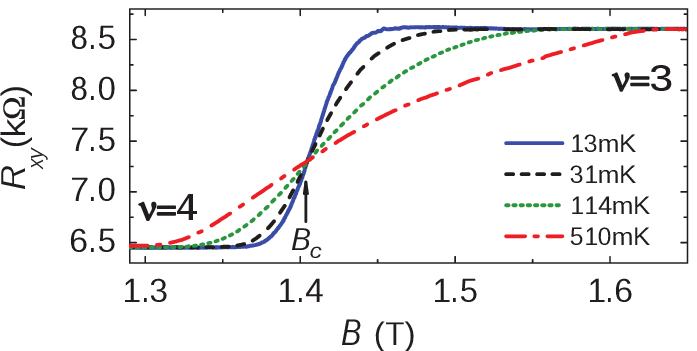}
 \caption{Hall resistance vs magnetic field for various temperatures, reproduced from \cite{QHEfinitesizeexp}. $B_c$ is the critical magnetic field where resistance doesn't change as temperature varies (colour online).}
\label{fig:plateautrans}}

The measurements in question are performed at the critical inter-plateaux magnetic field, $B_c$, defined as the field for which the Hall resistance remains constant as the temperature varies (see figure \ref{fig:plateautrans}). The quantity of interest is the slope of the resistance profile evaluated at the critical field, $(\partial R_{xy}/\partial B)_{B_c}$.

\FIGURE[!t]{
 \includegraphics[width=70mm]{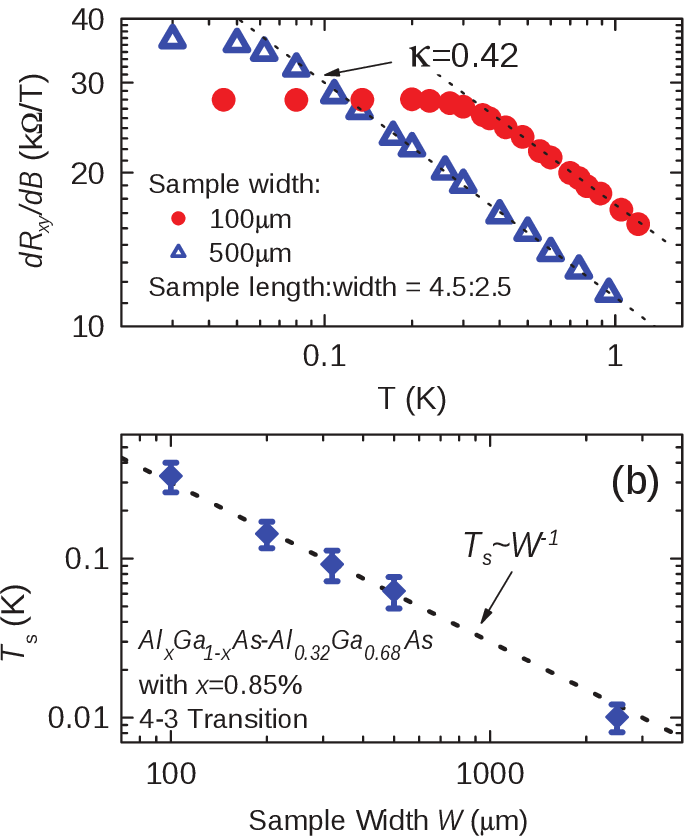}
 \caption{Top panel: $\partial R_{xy}/\partial B$ measured at $B = B_c$, as a function of temperature (reproduced from \cite{QHEfinitesizeexp}), showing the transition from power-law temperature dependence to temperature-independence. Bottom panel: The transition temperature between these regimes vs system size (colour online). \label{fig:finitesizeexp}}
}

This quantity is observed to be fairly sensitive to the amount of doping, $x$, in the Al$_x$Ga$_{1-x}$As/Al$_{0.32}$Ga$_{0.68}$As heterostructure, but for $0.6\% < x < 1.6\%$ a power-law behaviour
\begin{equation}
 \left( \frac{\partial R_{xy}}{\partial B} \right)_{B_c}
 \propto T^{-p} \,,
\end{equation}
is observed over two decades of temperature. Whenever this broad a range of scaling is seen the power is given by $p = 0.42 \pm 0.01$. This is seen in the top panel of figure \ref{fig:finitesizeexp}.

The significance of the sensitivity to doping is not yet clear, but deviations from $p = 0.42$ only arise when the power-law behaviour does not apply over as large a temperature range. The range of doping for which the robust scaling occurs reflects a regime where short-range scattering from the doped Al disorder dominates \cite{QHEfinitesizeexp,Alloy}, and further studies of this doping dependence could shed light on the domain over which the low-energy behaviour is universal, and so described by the CFT dual to our AdS description.\footnote{We thank Michael Hilke and Gabor Cs\'athy for helpful conversations on this point.}

Finite-size effects are observed when the derivative $(\partial R_{xy}/\partial B)_{B_c}$ is measured for samples small enough that the system size can compete with temperature effects in charge transport. The experiments find that at small enough system size, there exists a temperature, $T_s$, below which
$(\partial R_{xy}/\partial B)_{B_c}$ becomes independent of temperature. By repeating the measurements for samples of different size (see the bottom panel in figure \ref{fig:finitesizeexp}), it is found that
the transition temperature varies inversely with system size,
\begin{equation} \label{eqTsvsL}
 T_s \propto {1}/{L} \,,
\end{equation}
where $L$ is the sample width.

This relation has an interpretation in terms of critical scaling exponents \cite{ScalingRev}. At a critical point coherence length, $\xi$, and temperature are related by a power law, $\xi \propto T^{-s/2}$. If we follow ref.~\cite{QHEfinitesizeexp} and assume the coherence length saturates at the system size, $\xi \propto L$, for low enough temperatures, then eq.~\pref{eqTsvsL} implies $s=2$. This can be related to the dynamical exponent $z$ using relations among critical exponents. In particular, critical scaling theory implies $p = s/(2\nu)$, where $\nu$ is the exponent for the localization length as a function of the magnetic field: $\xi \propto \left| B-B_c \right|^{-\nu}$, as well as the scaling relation $z = 1/(p \nu)$. Combining these gives $z = 2/s$, which with $s = 2$ gives $z = 1$.

It is the apparent mismatch between this conclusion and the use of $z = 5$ of the model in \cite{Hallography} that we set out to understand in the next sections.

\section{Finite-size effects in AdS/CFT systems}
\label{sec:finitesizeADSCFT}

In this section we describe how finite-size effects are included into the holographic system of interest. Because the size introduces a new scale, $L$, this can be combined into a scale-invariant combination with temperature, $T$, allowing physical quantities to depend on these variables in a complicated way, even for scale invariant systems (like CFTs).
In \S\ref{sec-adsfinite} we first briefly review standard arguments about how finite-size effects can be used to describe a transition \cite{Witten-gauge} for AdS/CFT systems in the absence of a chemical potential. We then describe in \S\ref{sec-DBI-finite} how this argument generalizes to the AdS/QHE case, where a chemical potential plays an important role.

\subsection{Finite size with no chemical potential} \label{sec-adsfinite}

In this section we briefly summarize the discussion of ref.~\cite{Witten-gauge}, which discusses how to include finite-size effects for the simplest AdS/CFT system, and how these can change the low temperature properties of the system.
Following \cite{Witten-gauge}, consider the following bulk action
\begin{equation}
 S=-\frac1{2\kappa^2}\int\exd^4x\sqrt{-g}
 \left(R-\frac{6}{L^2}\right) \,,
\end{equation}
whose equations are solved by the black-hole metric
\begin{equation}
  \exd s^2 = - h(\tilde r) \, \exd t^2 + \frac{\exd {\tilde{r}}^2}{h(\tilde r)}
  + \tilde{r}^2\exd\Omega^2 \,, \label{adss2bh}
\end{equation}
with $\exd \Omega^2 = \exd \theta^2 + \sin^2 \theta \, \exd \phi^2$ and
\begin{equation}
  h(\tilde r) = \frac{\tilde{r}^2}{L^2}
  -\frac{\tilde{r}_h^3}{\tilde{r}L^2}
  +\left(1-\frac{\tilde{r}_h}{\tilde{r}}\right) \,.
\end{equation}
Here the integration constant is chosen so that $h(\tilde{r}_h) = 0$.

This differs from the translation-invariant black-brane solutions because the asymptotic, large-$\tilde r$, geometry is a sphere rather than a plane. Because the radius of the sphere provides a scale, this asymptotic geometry only becomes scale invariant at large $\tilde r/L$. Consequently the AdS radius, $L$, enters differently into observables, and ultimately plays the role of the CFT system size (as we now clarify).

\subsubsection*{CFT system size} \label{sec-systemsize}

Let us be a little more precise about what we mean by the system size for the CFT dual to the black-hole geometry, eq.~(\ref{adss2bh}). To this end consider the limit of large $\tilde r$ in (\ref{adss2bh}), for which the metric along slices of constant $\tilde r$ is
\begin{equation}
  \exd s^2_r \simeq - \frac{\tilde r^2}{L^2} \, \exd t^2
  + \tilde r^2 \, \exd\Omega^2
  = \frac{\tilde r^2}{L^2} \Bigl( - \exd t^2 + L^2
  \, \exd \Omega^2 \Bigr) \,.
\end{equation}
$\exd s^2_r$ is not quite the metric of the CFT, since the circumference of the time circle must be $1/T$ for a CFT at temperature $T$. Rather, $\exd s^2_r$ is conformal to the CFT metric, which is given by
\begin{equation}
   \exd s^2_\CFT = - \exd t^2+ L^2 \exd\Omega^2 \,.
\end{equation}
Clearly this metric describes a finite-size system with a real-space circumference of $2\pi L$, giving the precise relationship between system size and the AdS scale. Notice how this argument is independent of the small-$r$ geometry, so as long as the system is asymptotically AdS.

\subsubsection*{Temperature and system size}

An important consequence of having a finite system size can be seen from the expression for the Hawking temperature for this metric,
\begin{equation}
 4\pi T  = h'(\tilde r_h)
 = \frac{3\tilde{r}_h}{L^2} + \frac{1}{\tilde{r}_h} \,. \label{adss2temp}
\end{equation}
This expression is obtained, for example, by going to euclidean signature and requiring the time coordinate to be periodic, $t \simeq t + \beta$, with $\beta = 1/T$ chosen to remove the conical singularity that would otherwise appear at $\tilde r = \tilde r_h$.

What is noteworthy about eq.~\pref{adss2temp} is that it has a minimum at $\tilde{r}_h^2 = L^2/3$, with the temperature bounded below, $T \ge T_\star$, with
\begin{equation}
 4\pi T_\star = \frac{2\sqrt3}{L} \,.
\end{equation}
The system has a minimum temperature inversely proportional to the system size. But if the black hole does not describe temperatures lower than $T_\star$, what does? Ref.~\cite{Witten-gauge} proposes the system is better described by the alternative solution corresponding to $\tilde r_h = 0$, for which $h(\tilde r) = 1 + \tilde r^2/L^2$ does not vanish. In this case the periodicity, $\beta'$, of the euclidean time direction can be arbitrary (and so in particular can describe arbitrarily low temperatures).

To determine which solution the system prefers for any given temperature and system size we compare the two free energies, $F(T)$, for these solutions, using the AdS/CFT prescription that $F(T) = - TS_{\text{on-shell}}$. (Here $S_{\text{on-shell}}$ denotes the classical action evaluated at the classical solution, regarded as a function of its boundary values at large $\tilde r$.) Since the Einstein equation implies $R = 12/L^2$ for both solutions we have
\be
 F_{\beta'}(T) = \frac{3T}{\kappa^2 L^2} \int_0^{\beta'}
 \exd t \int_0^{\tilde{r}_\infty} \exd \tilde{r}
 \int\exd^2\Omega \, \tilde{r}^2 = \frac{8\pi T \beta' \tilde r_\infty^3 }{\kappa^2 L^2}\,,
\ee
and
\be
 F_{\beta}(T) = \frac{3T}{\kappa^2 L^2} \int_0^{\beta}
 \exd t\int_{\tilde{r}_h}^{\tilde{r}_\infty}
 \exd \tilde{r} \int\exd^2\Omega \, \tilde{r}^2
 = \frac{8\pi T \beta \, (\tilde r_\infty^3 - \tilde r_h^3)
 }{\kappa^2 L^2}\,,
\ee
with $\tilde{r}_\infty$ being a temporary regulator that is ultimately taken to infinity.

The parameters $\beta$ and $\beta'$ are related to one another in an $\tilde r_\infty$-dependent way, by the condition that both actions describe the same temperature, $T$, since this requires their euclidean time directions must have the same circumference for large $\tilde r$. That is, using the form of the two metrics at $\tilde r = \tilde r_\infty$,
\be
 \beta'\sqrt{(\tilde{r}_\infty)^2/L^2+1}= \beta\sqrt{(\tilde{r}_\infty)^2/L^2-r_h^3/(\tilde{r}_\infty L^2)+\left(1-\tilde{r}_h/\tilde{r}_\infty\right)}\nn
\ee
to give us the same CFT temperature at infinity, (that is, the $S^1$'s from the time component have the same circumference.)
Using this in the free-energy expressions gives
\bea
 \Delta F := F_{\beta}(T)-F_{\beta'}(T) &=& \frac{4\pi T}{\kappa^2 L^2}
 \Bigl[(\beta - \beta') \, \tilde{r}_\infty^3 - \beta \, \tilde{r}_h^3 \Bigr] \nn\\
 &=& \frac{4\pi\beta T}{\kappa^2 L^2} \left[ \tilde{r}_\infty^3
 \left(1 - \sqrt{ 1 - \tfrac{\tilde{r}_h (r_h^2/L^2+1)}{
 \tilde{r}_\infty (\tilde{r}_\infty^2/L^2+1)}}
 \right) - \tilde{r}_h^3\right] \nn\\
 &=&\frac{2\pi \tilde r_h}{\kappa^2} \left(
 \frac{L^2-\tilde{r}_h^2}{L^2} \right) \,,
\eea
where $\tilde r_\infty$ is taken to infinity in the last line.
This calculation indicates a transition at $\tilde{r}_h=L$, with the black-hole solution having lower free energy at larger $\tilde{r}_h$ (or higher temperatures), while the other solution has lower free energy for smaller temperatures. Notice in particular that this transition happens above the minimum temperature of the black-hole solution, which occurs at $\tilde{r}_h= L/\sqrt3$.

The value of the transition temperature can be seen by writing $\Delta F$ in terms of $T$, using \eqref{adss2temp}:
\be
 \Delta F=\frac{4\pi \tilde{r}_h}{\kappa^2}
 \left[1-\frac{L^2}{36} \left(4\pi T+\sqrt{(4\pi T)^2-\frac{12}{L^2}} \right)^2\right] \,,
\ee
which shows that $\Delta F = 0$ occurs at a transition temperature inversely proportional to the system size: $T_s \propto 1/L$.

\subsection{Finite size AdS/QHE black holes} \label{sec-DBI-finite}

We now repeat the above finite-size analysis for the quantum Hall-ography model. This involves several steps: construct the black hole solutions for asymptotic spherical geometries and see whether they have a minimum temperature; and if so, identify a candidate alternative gravity dual that describes the low-temperature phase. Once a low-temperature description is found we check whether its conductivity is temperature independent, and find how the temperature, $T_s$, changes with system size.

\subsubsection*{Field equations}

We start with the Hall-ography action of \S\ref{sec:qhallography}, and use the $SL(2,R)$ freedom to set $\chi = B = 0$, so
\be
 S_{\rm grav} + S_{\rm matter} = - \int \exd^4x \,
 \sqrt{-g} \; \left[ \frac{1}{2 \kappa^2}
 \left( R - \frac6{L^2} + \frac12 \, \partial_\mu\tilde \phi
 \, \partial^\mu\tilde \phi \right)
 + \mathcal T \Bigl( X - 1 \Bigr) \right] \,, \label{fulldbiaction}
\ee
and seek black hole solutions with spherical geometry at fixed radius and time. Adopting a dimensionless radial coordinate, %
\be \label{dimlessrdef}
 \tilde r = L \, r \,,
\ee
and using the metric ansatz
\be \label{metricansatz}
 \exd s^2 = \left[ -h(r) \, e^{\xi(r)} \exd t^2
 + \frac{L^2\exd r^2}{h(r)} + L^2 r^2 \exd\Omega^2 \right]
 \quad \hbox{and} \quad
 \tilde\phi = \tilde\phi(r)  \,,
\ee
we find (as above) that it is $L$ that plays the role of the CFT system size, with the CFT metric conformal to the asymptotic bulk metric.
The temperature associated with this metric is
\be
4\pi T = \frac1L \, e^{\xi(r_h)/2} \, h'(r_h) \,, \label{generaltemp}
\ee
where $h(r_h) = 0$.

The Maxwell field equation integrates simply to give
\be
 \\
 G^{rt} = \frac{\mathcal T\ell^4}{X} \,
 e^{-\phi} F^{rt} = \frac{Q \, e^{-\xi/2}}{L^3 r^2} \label{Qdef}
\ee
where $Q$ is the black hole electric charge. It is shown in \cite{Hallography} that the CFT current density is given by $J^i = \sqrt{-g} \; G^{ri}|_\infty$, and keeping in mind that $g_{\mu\nu}$ is only asymptotically conformal to the CFT metric in the finite-volume case, the CFT charge density turns out to be related to $Q$ by
\begin{equation}
 Q = L^2 \rho_{\CFT} \,.  \label{charge-den-rel}
\end{equation}
Consequently $Q$ scales like $L^2$ if $L$ is varied with fixed CFT charge density. Since this expression also shows that $Q$ counts the total number of charge carriers in the CFT, $Q \gg 1$ is the regime appropriate to real quantum Hall systems. How finite size affects the conductivity calculation for this system is explored in appendix \ref{sec-condcalc}.

The field equations differ slightly from the black brane case solved in \cite{Hallography}, because of the curvature of the surfaces of fixed $r$ and $t$, which appears in the $\theta\theta$ Einstein equation. This difference is less and less important for larger $r$. The field equations to solve are
\be \label{finitefieldeqxi}
 r\xi'+\frac12(r \phi')^2 = 0 \,,
\ee
and
\bea \label{finitefieldeq}
 \frac{h'}{r} + \frac{h\xi'}{2r} + \frac{h}{r^2}
 -\frac{1}{r^2} - 3 - \mathcal{\hat T}\left(
 \frac{X-1}{X} \right) &=& 0 \\
 \frac{e^{-\xi/2}}{r^2} \left( e^{\xi/2} \,r^2 h \, \phi'\right)' +\frac{\mathcal{\hat T}}{X}
 \left( X^2 - 1 \right) &=& 0 \,,\nn
\eea
where $r$ is the dimensionless radial coordinate, and the other quantities in these equations are
\be \label{Xh nearhorizon form}
 X^{-2} = 1 + \frac{Q^2 e^{\phi}}{\mathcal{\hat T}^2 r^4} \,,
\ee
$\mathcal{\hat T} = \kappa^2 L^2 \mathcal T$ and  $\phi = \tilde\phi - 4\log(\ell/\kappa)$.

\subsubsection*{Asymptopia: the far-field $z=1$ region} \label{sec-z1}

The first region of interest is the asymptotic far-field regime, for which $r \gg 1$. An approximate solution in this regime is obtained by expanding fields in powers of $1/r$ plugging the result into \eqref{finitefieldeq}. The approximate solution obtained in this way is $\xi(r) \simeq 0$,
\bea \label{z1soln}
 h(r) &\simeq& r^2 \left( 1 + \frac1{r^2} - \frac{r_b^3}{r^3}
 + \frac{Q^2e^{\phi_0}}{2\mathcal{\hat T}r^4} \right)\nn\\
 &=& r^2 \left[ 1 - \frac{r_h^3}{r^3}
 + \left( \frac1{r^2} - \frac{r_h}{r^3} \right)
 + \frac{Q^2e^{\phi_0}}{2\mathcal{\hat T}} \left(
 \frac1{r^4} - \frac1{r_hr^3} \right) \right] \nn\\
 \hbox{and} \quad
 \phi &\simeq& \phi_0 + \frac{\phi_1}{r^3} +
 \frac{Q^2e^{\phi_0}}{4\mathcal{\hat T} r^4}
 \,,
\eea
where $\phi_0$ and $\phi_1$ are boundary data to be specified, and $r_b$ is an integration constant that is traded in the second line for $r_h$, defined as the position satisfying $h(r_h)=0$.

More specifically, this solution assumes that
\begin{equation} \label{z1approx}
 \frac1{r^2} \sim
 \frac{Q^2e^{\phi_0}}{\mathcal{\hat T}^2r^4} \sim
 \frac{\phi_1}{r^3} \ll 1 \,,
\end{equation}
so that anything quadratic in these quantities can be neglected. But we can only use eq.~\pref{z1soln} to infer that $h(r)$ vanishes for some $r$ if $r_h$ is large enough to ensure that eqs.~\pref{z1approx} is valid for all $r_h < r < \infty$. When this is true a horizon exists in this region, terminating the solution at $r = r_h$. Using equation \eqref{generaltemp} the temperature for this metric becomes
\be
  4\pi T = \frac1L \left(2r_h + \frac1{r_h}-\frac{Q^2e^{\phi_0}}{2\mathcal{\hat T}}\right).
\ee

\subsubsection*{Attractor: the near-horizon $z=5$ region} \label{sec-z5}

A different solution describes the near-horizon geometry if
${Q^2e^{\phi_0}}/{\mathcal{\hat T}^2 r_h^4}$ is not much smaller than unity, and provided $r_h$ is still large enough to be in the large-$r$ limit. Then what was an exact solution for the black brane in ref.~\cite{Hallography},
\be
 h(r) = h_0 r^2 \left( 1 - \frac{r_h^7}{r^7} \right)
 \,, \quad
 \xi(r) = 8 \ln\left( \frac r{r_c} \right)
 \quad \hbox{and} \quad
 \phi(r) = 4 \ln\left( \frac r{r_c} \right) + \phi_0 \label{z5soln} \,,
\ee
also solves eqs.~\eqref{finitefieldeq}, provided we drop terms of relative order $r^{-2}$. In these expressions $r_c$ is an integration constant, which is ultimately fixed by matching to the large-$r$ $z=1$ region below ({\em c.f.} eq.~\eqref{hightenattractedvalue}).

\FIGURE[m]{
 \centering
 \includegraphics[width=60mm]{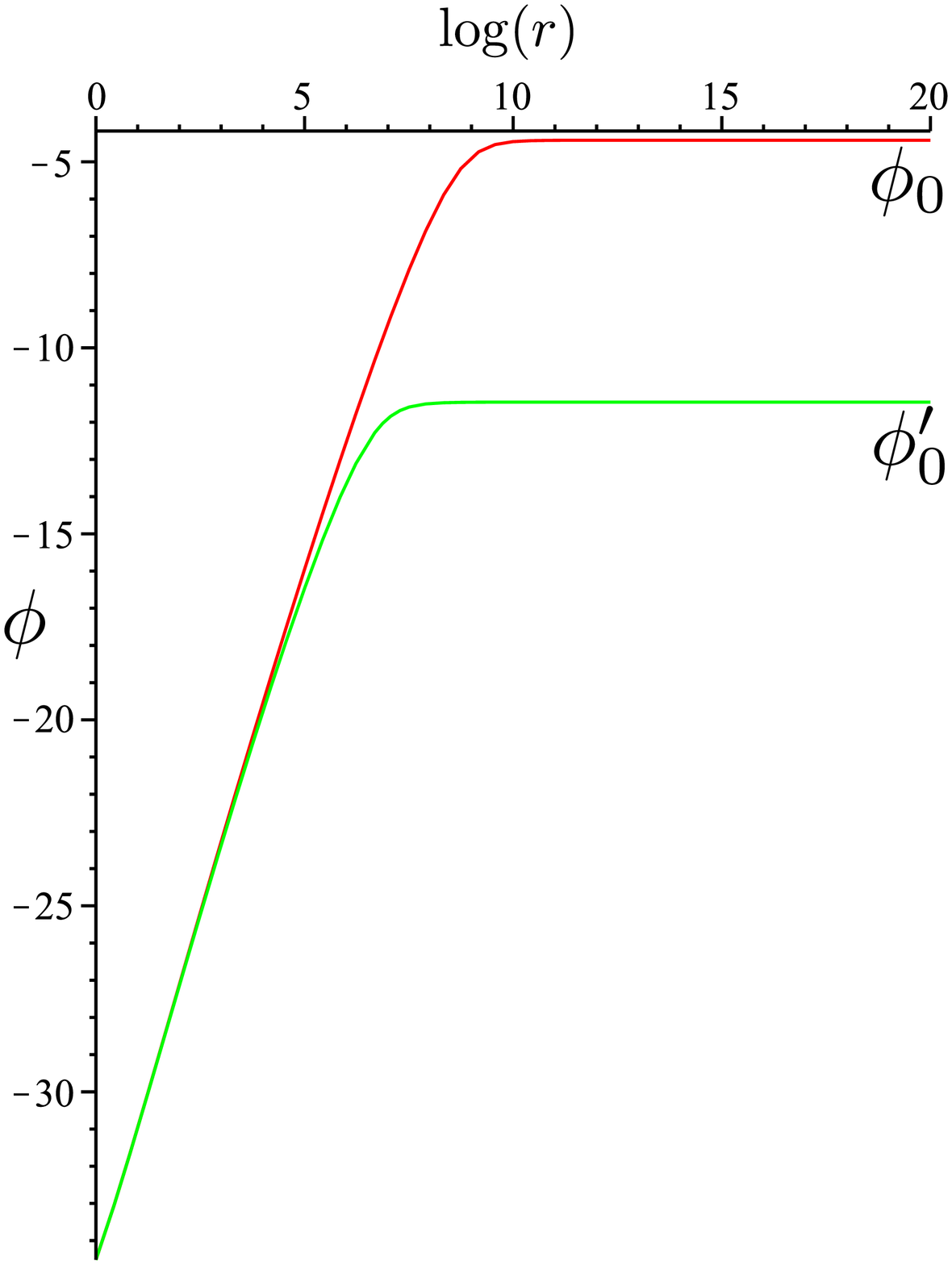}
 \caption{Semi-log plot of the dilaton profile for $Q=1000$ and $\mathcal{\hat T}=10^{-5}$, for two asymptotic initial conditions (colour online). \label{fig:dilatoncrossover}}}

For this solution the quantity $X$ evaluates to a constant,
\be
 X := X_h = \frac{-(6+2\mathcal{\hat T}) + \sqrt{(6+2\mathcal{\hat T})^2 + 5\mathcal{\hat T}^2}}{ \mathcal{\hat T}} \,,
\ee
in terms of which $h_0$ is fixed by the equations of motion
\be
 7 \, h_0 = 3 + \mathcal{\hat T} \left( \frac{X_h-1}{X_h} \right) \,.
\ee
Notice that $X_h \to 1 + \cO(1/\mathcal{\hat T})$ as $\mathcal{\hat T} \to \infty$ and $X_h \to \frac{5}{12} \, \mathcal{\hat T} + \cO(\mathcal{\hat T}^2)$ as $\mathcal{\hat T} \to 0$.

As discussed\footnote{Although ref.~\cite{Shamit-attractor} uses the Maxwell action, ref.~\cite{Hallography} shows their attractor solution also applies for black branes with the DBI action, and so also in the present case for large enough black holes.} in ref.~\cite{Shamit-attractor}, this solution is an attractor inasmuch as all solutions eventually approach this one in the near-horizon limit, regardless of their boundary conditions at large $r$. We have checked numerically that these solutions exist, and fig.~\ref{fig:dilatoncrossover} shows two examples where $\phi(r)$ asymptotes to eqs.~\pref{z1soln} for very large $r$, but then crosses over to the attractor form, eq.~\pref{z5soln}, as the horizon is approached. The figure also shows how the solution for $\phi(r)$ crosses quite quickly from the logarithmic behaviour of eq.~\pref{z5soln} to the asymptotically constant limit of eq.~\pref{z1soln} (similar to what was also found in ref.~\cite{Shamit-attractor}).

In what follows it is crucial to determine how the integration constant $r_c$ depends on system parameters like $L$ and $\rho_\CFT$, a dependence that arises when the attractor solution is matched onto the solution at asymptotically large $r$. To this end it is useful to treat the transition between \pref{z1soln} to \pref{z5soln} as occurring at a specific transition radius, since fig.~\ref{fig:dilatoncrossover} shows this to be a good approximation. Since $\xi = 0$ in the asymptotic region, matching $\xi(r)$ using eq.~\pref{z5soln}, shows this transition radius is given by $r = r_c$. The dependence of $r_c$ on other parameters is then obtained by similarly demanding continuity of $\phi(r)$ at $r = r_c$, leading to
\bea
  \frac{Q^2 \, e^{\phi_0}}{ r_c^4} =
  \frac{\mathcal{\hat T}^2 - X^2_h(\mathcal{\hat T})}{ X_h^2(\mathcal{\hat T})}
  &\simeq& \mathcal{\hat T}^2 + \cO(\mathcal{\hat T})
  \qquad\qquad \hbox{if $\mathcal{\hat T} \gg 1$} \label{hightenattractedvalue} \\
  &\simeq& \frac{119}{25} \Bigl[ 1 + \cO(\mathcal{\hat T}) \Bigr] \qquad \hbox{if $\mathcal{\hat T} \ll 1$} \,.\nn
\eea
Notice in particular that no choice for $\mathcal{\hat T} > 0$ allows $Q \, e^{\phi_0/2}/r_c^2 \ll 1$.

Eliminating $Q$ using $Q = \rho_\CFT L^2$ and using the definition $\hat \cT = \kappa^2 L^2 \cT$ then gives
\bea \label{rcvsLformulae}
 r_c^4 &\simeq& \frac{\rho_\CFT^2 e^{\phi_0}}{\kappa^4 \cT^2} \qquad\qquad\quad \hbox{if $\mathcal{\hat T} \gg 1$} \nn\\
 &\simeq& \frac{25}{119} \, \rho_\CFT^2 e^{\phi_0} L^4 \qquad\; \hbox{if $\mathcal{\hat T} \ll 1$}
\eea
which shows that when $L$ is varied with fixed $\phi_0$ and $\rho_\CFT$, $r_c$ is independent of $L$ for large $\cT$ but scales as $L^4$ when $\cT$ is small. We see that the position of the cross-over between the $z=1$ and $z=5$ asymptotic solutions becomes $L$-dependent when the dimensionful parameter $\cT$ is sufficiently large.

This $L$-dependence of $r_c$ is of interest because it enters into the relation between the temperature and $r_h$. Using eq.~\eqref{generaltemp} as before to connect $T$ and $r_h$, using the $z=5$ attractor geometry gives
\be
 4\pi T = \frac{7h_0 r_h^5}{L r_c^4}, \label{z5temp}
\ee
where the $r_c$ dependence arises from the form of $\xi(r)$ in \eqref{z5soln}. Finally, using eqs.~\pref{rcvsLformulae} to eliminate $r_c$ in \eqref{z5temp} gives a form which makes the scaling of the $T-r_h$ relation with system size more explicit:
\bea \label{z5Tcrossoverdep}
 4\pi T &=& \left( \frac{\kappa^4\mathcal T^2 r_h^5}{ \rho_\CFT^2e^{\phi_0}}\right) \frac{7h_0}{L}  \qquad\qquad\qquad\hbox{if $\mathcal{\hat T} \gg 1$} \nn\\
 4\pi T &=& \left(  \frac{119}{25} \frac{r_h^5}{\rho_\CFT^2e^{\phi_0}} \right) \frac{7h_0}{L^5} \qquad\qquad \hbox{if $\mathcal{\hat T} \ll 1$} \,.
\eea

This last pair of equations has two important consequences if we anticipate two results from subsequent sections.

\medskip \noindent {\em 1. Temperature scaling:}

\smallskip \noindent
First, as is shown in Appendix \ref{sec-condcalc} (and ref.~\cite{Hallography}), the power-law temperature dependence of the Hall conductivity (and resistivity) arises because it scales proportional to $1/r_h^2$. Whenever $r_c > r_h$ eq.~\pref{z5Tcrossoverdep} then implies a temperature scaling proportional to $T^{-2/5}$, for {\em both} large and small $\mathcal{\hat T}$.

What is required to ensure\footnote{Recall the condition $r_h \gg 1$ is required to use the large-$r$ black-hole solution described above. (The discussion of later sections examines what happens if this condition is relaxed.)} $r_c > r_h \gg 1$? For $\mathcal{\hat T}$ large inspection of eq.~\pref{hightenattractedvalue} shows this requires $1 \ll r_h^2 \ll {Q \, e^{\phi_0/2}}/{\mathcal{\hat T}}$, which is a non-empty region only if $\mathcal{\hat T} \ll Q \, e^{\phi_0/2}$. In particular if $\mathcal{\hat T} \gg 1$ a large-$r$ crossover between the two solutions requires $Q$ must be even larger, since the use of semi-classical reasoning requires $e^{\phi_0} \ll 1$. Happily, as mentioned earlier, large $Q$ is the regime of interest for real quantum Hall systems.

\medskip \noindent {\em 2. Scaling with System Size:}

\smallskip \noindent
The second important consequence of eq.~\pref{z5Tcrossoverdep} is what it says about how $T$ scales with $L$ when $r_h$ is fixed. Fixed $r_h$ is of interest because this is what subsequent sections show is required for the transition temperature, $T_s$.

Eq.~\pref{z5Tcrossoverdep} shows that if $\mathcal{\hat T}$ is negligible then $T$ varies with $L$ (for fixed $r_h$) as expected for $z=5$ scaling: $T \propto 1/L^5$. In this case the only scales that arise are those characterizing the CFT, such as $\rho_\CFT$, $T$ and $L$. However, the same is not true when $\mathcal{\hat T}$ is large due to the presence of the additional scale $\cT$. In this limit the $L$-dependence of $r_c$ ensures we instead have $T \propto 1/L$, provided $r_h$ is held fixed.

\subsubsection*{Relaxing the large-$r$ approximation} \label{sec-perturbfe}

This section describes how the black hole temperature is related to $r_h$ and other system parameters in the regime where the above approximate forms need not apply because we no longer neglect the $1/r^2$ term in \eqref{finitefieldeq}; {\em i.e.}~where the black hole solutions can differ from the black-brane solutions of ref.~\cite{Hallography}. Our goal is to motivate the existence of a transition to a new regime by showing that $T(r_h)$ is bounded from below, with a minimum temperature, $T_\star$, below which some other solution must replace the black-hole description.

Although we don't have the luxury of an explicit solution in the general case, because our main interest is in how the temperature depends on other parameters like $r_h$ we can proceed by expanding the metric in powers of $r - r_h$, as follows
\bea
 h(r) &=& h_1 (r-r_h)+ h_2(r-r_h)^2 + \cdots\nn\\
 \phi(r) &=& \phi_h + \frac{\omega}{r_h} \, (r-r_h) + \cdots \label{smallrhansatz}\\
 \xi(r) &=& \xi_h - \frac{\omega^2}{2r_h} \, (r-r_h) + \cdots \,, \nn
\eea
where we define $\phi_1 := \omega/r_h$ and eq.~\pref{finitefieldeqxi} is used to relate the linear terms in $\phi$ and $\xi$ to one another. In the special case where $r_h \gg 1$ the solutions of the previous sections apply, with the solutions of eqs.~\pref{z1soln} corresponding to the choices $\omega \simeq \xi_h \simeq 0$ and $\phi_h \simeq \phi_0$, while those of eqs.~\pref{z5soln} being captured by $\omega \simeq 4$, $\phi_h \simeq \phi_0 + 4 \ln(r_h/r_c)$ and $\xi_h \simeq 8 \ln(r_h/r_c)$.

Since eq.~\pref{finitefieldeqxi} has been used already to relate $\xi$ to $\phi$, there are only two other field equations to solve. Expanding equations \eqref{finitefieldeq} about $r=r_h$ gives
\bea
 &&\frac{h_1}{r_h} - 3 - \frac1{r_h^2}
 + \frac{\mathcal{\hat T}}{X_h}(1-X_h) \\
 && \qquad\qquad + \left[
 \frac{2h_2}{r_h} + \frac2{r_h^3} + \frac{\omega^2h_1}{4r_h^2} - \frac{(4-\omega) Q^2 \, e^{\phi_h} X_h}{2 r_h^{5-\omega} \mathcal{\hat T}} \right](r-r_h) + \cdots = 0 \,,\nn \label{feexpansion1}
\eea
and
\bea
 &&\frac{\omega h_1}{r_h} + \frac{\mathcal{\hat T}}{X_h}
 (1-X_h^2) \\
 && \qquad\qquad + \left[ \frac{2\omega h_2}{r_h}
 +\frac{h_1 \omega^3}{4r_h^2} + \frac{(4-\omega) Q^2
 e^{\phi_h}}{ r_h^{5-\omega}\mathcal{\hat T}} \, X_h\left(1+X_h^2\right) \right](r-r_h) + \cdots = 0 \,,\nn
  \label{feexpansion2}
\eea
where
\be
 X_h^{-2} = 1+\frac{Q^2e^{\phi_h}}{\mathcal{\hat T}^2 r_h^{4-\omega}} \,.
\ee
Since these equations hold for all $r$ satisfying $|r-r_h| \ll 1$, the coefficient of each power of $r-r_h$ must separately vanish. Working to linear order in $r-r_h$ then implies four conditions, which we solve for $h_1, \omega, h_2$, and $\phi_h$ as functions of $r_h$. The left panel of figure \ref{fig:h1omegaTTprhdep} plots sample numerical solutions for $h_1$ and $\omega$ obtained in this way, as functions of $r_h$. In principle this can be continued to higher orders in $(r-r_h)$, though the higher terms have no bearing on how the temperature of the system depends on $r_h$.

\FIGURE[t]{
 \centering
 \includegraphics[width=60mm]{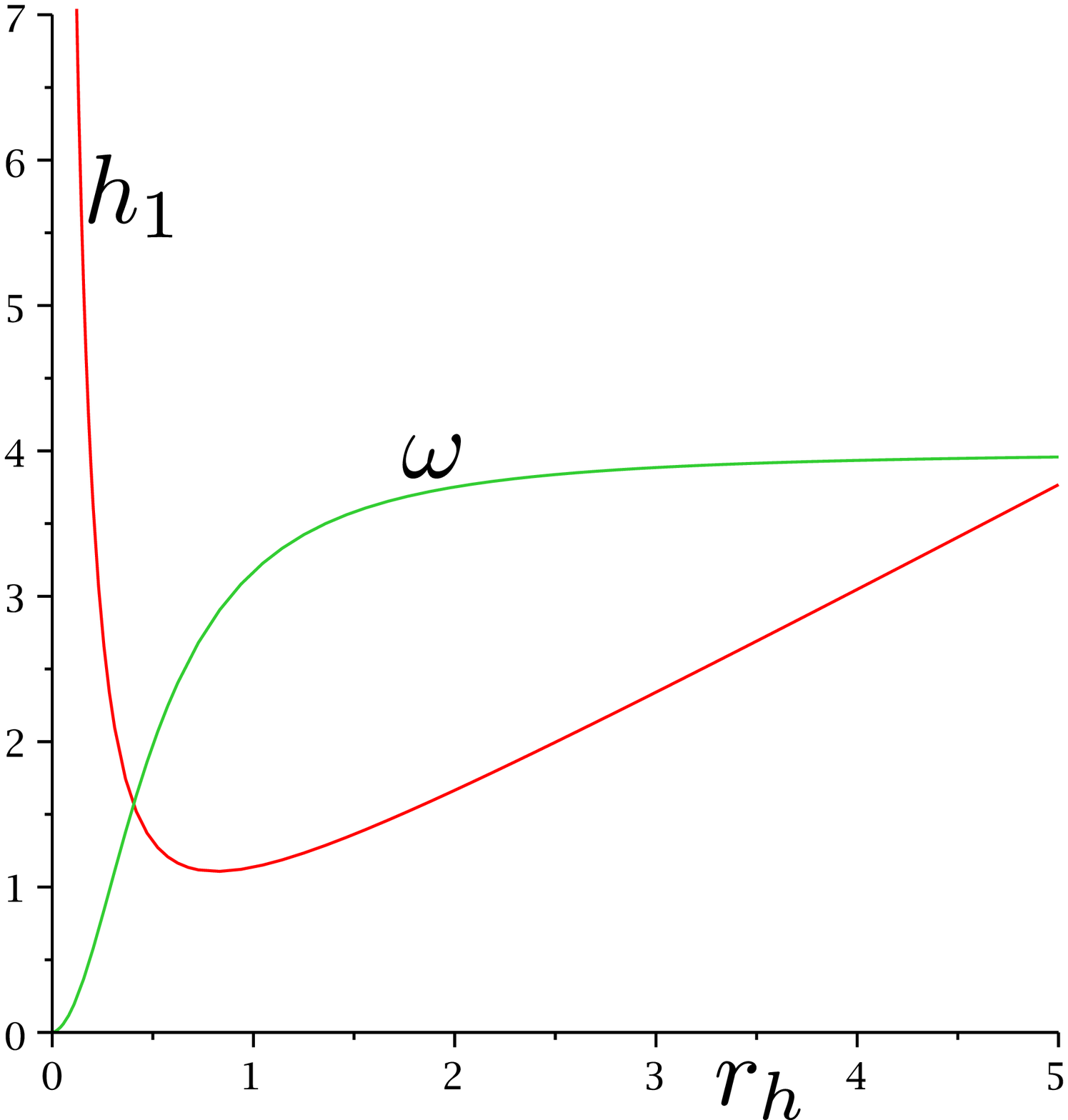}
 \includegraphics[width=60mm]{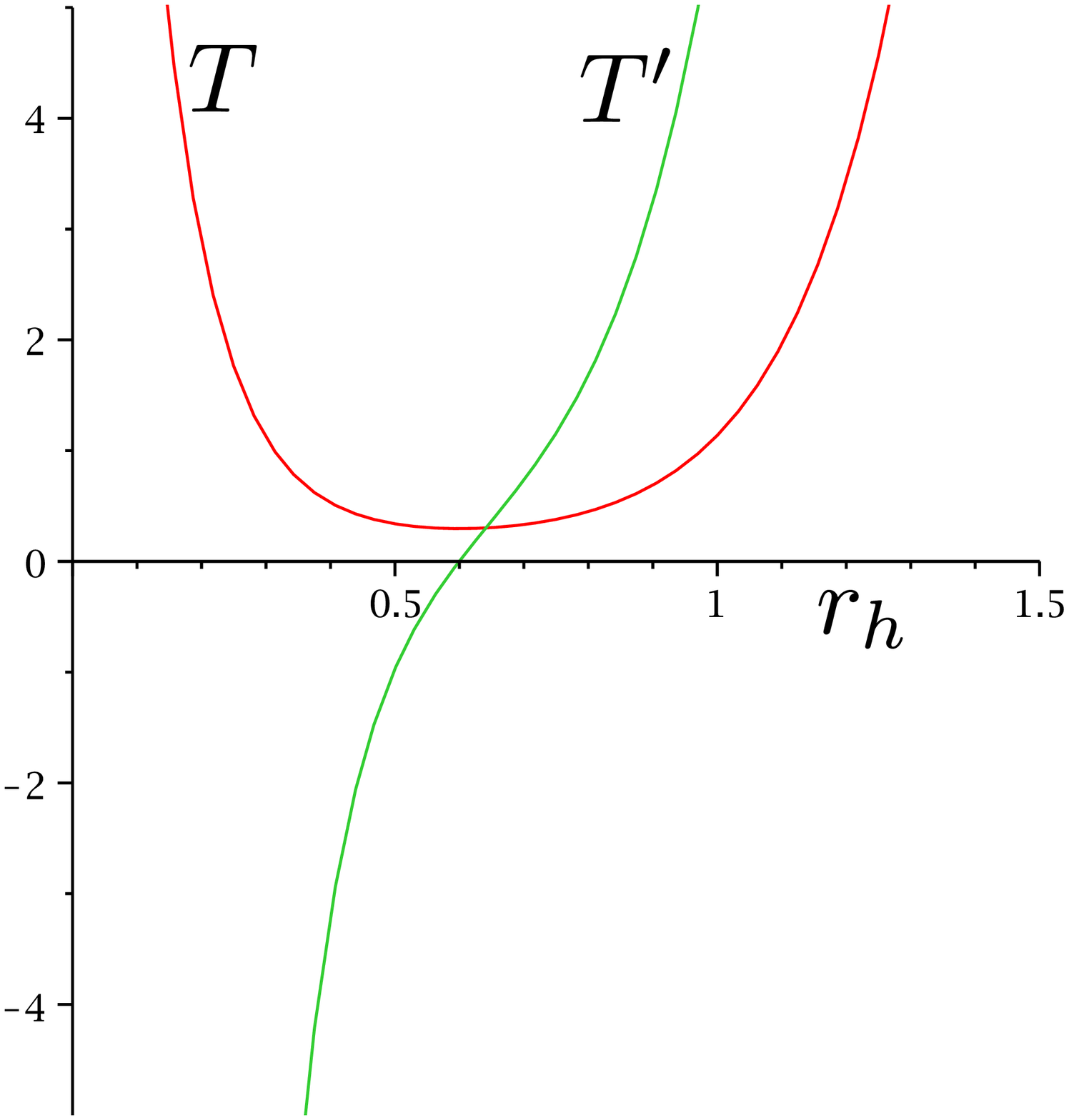}
 \caption{Left panel: A plot of $h_1$ and $\omega$ as functions of $r_h$.  Note that $h_1$ diverges and $\omega \rightarrow 0$ as $r_h\rightarrow0$, a reflection of the minimum temperature discussed in the main text. Also note that $\omega \to 4$ and $h_1 \to r_h$ at large $r_h$, in agreement with the analytic large-$r_h$ solution in the $z=5$ region.  Right panel: A plot of $T(r_h)$ and $\exd T/\exd r_h (r_h)$ against $r_h$ showing that temperature has a minimum for the black hole solution. In both cases we choose the value $\mathcal{\hat T} = 1$ (colour online). \label{fig:h1omegaTTprhdep}}
}

While these equations don't all have analytic solutions, the equation for $h_1$ is a quartic equation and so does have a closed-form solution in principle. This allows a check on our numeric solutions, since one can compare with the known $z=1$ and $z=5$ solutions found earlier by taking the large-$r_h$ limit. To obtain $z=1$ take $\mathcal{\hat T} \rightarrow \infty$, to find
\be
 h_1 = 3r_h + \frac1{r_h} \quad \hbox{and} \quad
\omega = 0 \,,
\ee
agreeing with eqs.~\eqref{z1soln}. Similarly, to get the $z=5$ solution take $r_h\gg1$ while ensuring that $Q^2 e^{\phi_0}/ \mathcal{\hat T}^2 \gsim r_h^4$, in which case $\omega=4$ and the value found for $h_1$ agrees with \eqref{z5soln} (as is also seen from the left panel of fig.~\ref{fig:h1omegaTTprhdep}).

In terms of these quantities the temperature for the system is given by eq.~\pref{generaltemp},
\begin{equation}
 4\pi T = e^{\xi_h/2} \, \frac{h_1}L \,, \label{smallrhtemp}
\end{equation}
where we keep in mind that this expression implicitly depends on the quantities $\mathcal{\hat T}$ and $Q$ through their appearance in $\xi_h$ and $h_1$. A plot of this expression for $T$ (and its derivative with respect to $r_h$) is shown in the right panel of figure \ref{fig:h1omegaTTprhdep}, which explicitly shows how $T$ has a minimum at a particular value $r_h = r_{h\star}$.

The upshot is that there is a minimum temperature the black hole can describe, and so some other solution must describe the physics below this temperature, similar to the discussion in section \S\ref{sec-adsfinite}. In the present instance this new solution cannot simply be anti-de Sitter space because of the existence of the black hole charge. Our proposal for this new solution is a `stellar' configuration, described in section \S\ref{sec-TOVequations} below.

\subsubsection*{$L$-dependence of the minimum temperature}

Before turning to what describes the very low-temperature regime, we first pause to examine how the minimum temperature, $T_\star$, varies with $L$ (when $\rho_\CFT$, $\phi_0$ and $\cT$ are fixed). We do so as a warm-up to a similar discussion for the transition temperature, $T_s$, between the high- and low-temperature regimes.

\FIGURE[ht]{
 \centering
 \includegraphics[width=60mm]{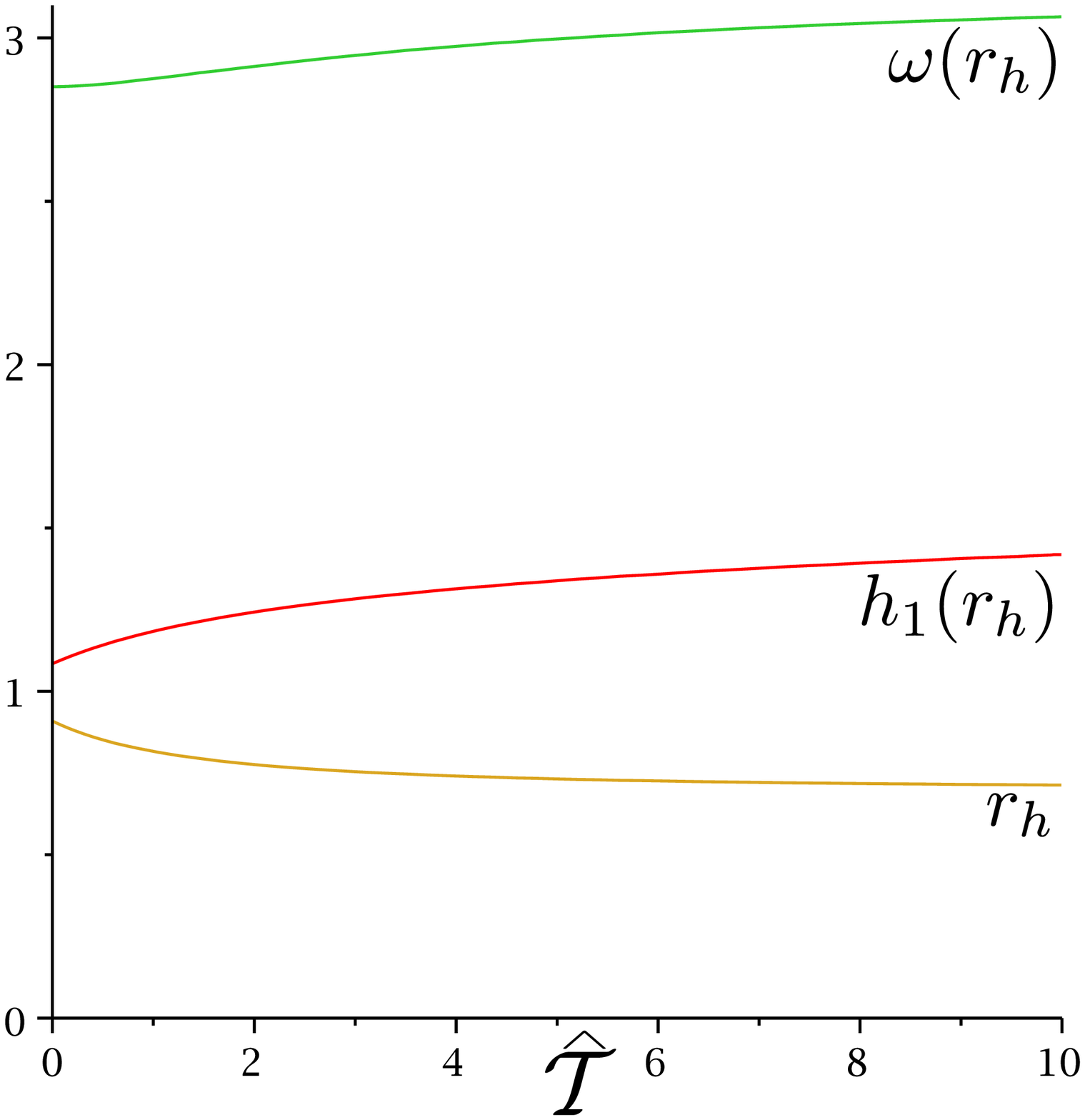}
 \caption{Plots, from top to bottom, of $\omega_\star$, $h_{1\star}$ and $r_{h\star}$ vs $\mathcal{\hat T} = \kappa^2 L^2 \mathcal T$, were `$\star$' denotes evaluation at the $r_h$ that minimizes the temperature $T(r_h)$. The weak dependence on $\mathcal{\hat T}$ $L$ dependence on implies that $T_{\star}$ is roughly proportional to $L^{-1}$ (colour online). \label{fig:h1rhomegatdep}}
}

Naively, eq.~\pref{smallrhtemp} suggests that $T_\star$ is inversely proportional to $L$, however this ignores the potential $L$-dependence that the parameters $\xi_{h\star}$ and $h_{1\star}$ might acquire due to their implicit dependence on the quantities $Q = \rho_\CFT L^2$ and $\mathcal{\hat T} = \kappa^2 L^2 \mathcal{T}$ appearing in eqs.~\pref{feexpansion1} and \pref{feexpansion2}, as well as through the dependence on $r_c$ that $\xi_h$ implicitly acquires once the solution is matched onto the solution at infinity.\footnote{See, for example, the discussion below eq.~\pref{smallrhansatz} showing $\xi_h \simeq 8 \ln(r_h/r_c)$ in the large-$r_h$ limit.}

The dependence of $r_{h\star}$, $h_{1\star}$, and $\omega_\star$ on $\mathcal{\hat T}$ is plotted in figure \ref{fig:h1rhomegatdep}, showing this dependence to be comparatively weak, and becoming weaker for larger $\mathcal{\hat T}$. This weak dependence on $\mathcal{\hat T}$ can be understood analytically as follows. The starting point observes that because $X = \cO(\mathcal{\hat T})$ for small $\mathcal{\hat T}$ and $X = 1 + \cO(1/\mathcal{\hat T})$ when $\mathcal{\hat T}$ is large, the quantity $\mathcal{\hat T} (1-X)/X$ is independent of $\mathcal{\hat T}$ in both of these limits. In this case, inspection of eqs.~\pref{feexpansion1} and \pref{feexpansion2} shows that the only other relevant $\mathcal{\hat T}$-dependence comes from the combination $Q^2 e^{\phi_h} X/\mathcal{\hat T}$, and so in particular the solutions for $h_1$, $\omega$, $h_2$ and $\phi_h$ must be independent of $\mathcal{\hat T}$ if $Q^2 e^{\phi_h} X/\mathcal{\hat T} \ll 1$ and $\mathcal{\hat T}$ is either much larger or smaller than 
unity.

Whenever this is true, the only remaining potential $L$-dependence in $T_\star$ is the implicit $L$-dependence appearing through $\xi_{h\star}$'s dependence on $r_c$. This $r_c$-dependence can be explicitly traced, such as is described in detail below eqs.~\pref{z5temp} for the large-$r_h$ limit (where $\xi$ is given by eq.~\pref{z5soln}). Because $T_\star$ is defined at fixed $r_h = r_{h\star}$, the arguments given below eqs.~\pref{z5temp} show that $T_\star \propto 1/L^5$ when $\mathcal{\hat T}$ is small, while $T_\star \propto 1/L$ when $\mathcal{\hat T}$ is large.

\section{The low-temperature phase: a charged star} \label{sec-TOVequations}

We now describe the alternative solution that describes the AdS/QHE system for small temperatures. This cannot be the empty AdS solution considered in ref.~\cite{Witten-gauge} because of the requirement that the solution carry electric charge $Q$. For this reason in this section we propose a charged star as the alternative low-temperature solution, and interpret the transition between the low-temperature and high-temperature phases of the CFT as corresponding on the gravity side to the transition, with increasing mass, from a stellar solution to a black hole solution.

This section develops this proposal in three steps. First \S\ref{sec:hydrostar} solves the equations of hydrostatic equilibrium to obtain the properties of a charged star. Then \S\ref{sec-phasediagram} compares the free energy of this solution with the corresponding black hole solutions, identifying in particular the temperature, $T_s$, at which the transition between a star and a black hole occurs. Finally, \S\ref{sec:curveshape} examines how $T_s$ changes with $L$.

\subsection{Hydrostatic equilibrium for a charged star}
\label{sec:hydrostar}

To find the new low-temperature solution we solve the Tolman-Oppenheimer-Volkoff (TOV) equations for a charged star in asymptotically AdS space. We ultimately solve these equations numerically, using values for the stellar mass and charge (and asymptotic dilaton) corresponding to those used in calculating the free energies for both the black hole and stellar solutions in later sections.

For the action of a charged perfect fluid in the stellar interior we follow \cite{Hartnoll:2010gu} and use
\be
 \mathcal L_{\text{fluid}} = - \sqrt{-g} \Bigl[ \rho_m
 + \lambda_1 (u^\mu u_\mu + 1 ) + \rho_c \,
 u^\mu (\nabla_\mu \lambda_2+ A_\mu) \Bigr] \,,
\ee
where $u^\mu$ is the fluid's 4-velocity, and $\rho_m$ and $\rho_c$ are is its mass and charge densities. $\lambda_1, \lambda_2$ are Lagrange multipliers, introduced to enforce the 4-velocity condition $u^\mu u_\mu = -1$ and conservation of charge $\nabla_\mu (\rho_c u^\mu) = 0$.

The stress-energy tensor for this action is
\begin{equation}
 T^f_{\mu\nu}=(\rho_m+p)u_\mu u_\nu +p \, g_{\mu\nu} \,,
\end{equation}
to which we add the energy-momentum tensor for the dilaton and the DBI action,
\begin{equation}
 T^\DBI_{\mu\nu}=\frac{2}{\sqrt{-g}} \left( \frac{\partial\mathcal L_{\DBI}}{\partial g^{\mu\nu}}
 \right) \,,
\end{equation}
to obtain the full energy momentum tensor for our DBI-charged star,
\bea
 T_{\mu\nu} &=& (\rho_m+p) u_\mu u_\nu
 + p \, g_{\mu\nu}
 - \frac{1}{4\kappa^2} \left[ g_{\mu\nu} \partial^\lambda \phi\partial_\lambda \phi
  -2 \partial_\mu\phi \partial_\nu\phi \right] \nn\\
  && \quad\quad - g_{\mu\nu} \mathcal T (X-1)
  + G_\mu^{\ \lambda}F_{\nu\lambda} \,.
\eea

Because $SL(2,R)$ invariance requires the stress tensor to be invariant, we imagine $\rho_m$, $p$ and $u^\mu$ to be $SL(2,R)$ invariant. We cannot also do so for the charge density, $\rho_c$, because we know that the Maxwell field (and in particular the total electric charge $Q$) transforms. For general transformations electric charge gets mapped into magnetic charge, and so specifying a fully $SL(2,R)$ invariant fluid would require having both electric and magnetic charge densities. We side-step this issue here, and for the purposes of comparing to black hole solutions with axion and magnetic fields turned off we consider only electrically charged matter and ignore the axion.

The Einstein equation is as before (but using this new stress energy tensor); the dilaton field equation remains unchanged; and the Maxwell equation inside the star is modified to
\begin{equation}
 \nabla_\mu G^{\mu\nu}=J^\nu=\rho_c u^\nu \,.
\end{equation}
The field equation describing the motion of the fluid is given by conservation of energy-momentum
\bea
 \nabla_\mu T^{\mu\nu} &=&
 -\frac{1}{4\kappa^2} \left[ 2(\partial_\lambda\phi)
 \nabla^\nu\nabla^\lambda\phi -2 ( \partial_\mu\phi )
 \nabla^\mu \nabla^\nu \phi - 2 ( \partial^\nu\phi )
 \Box\phi\right] + \nabla_\mu \left[ u^\mu u^\nu ( \rho_m + p ) + p \, g^{\mu\nu} \right] \nn\\
 && \qquad\qquad - \mathcal T \left[ \frac{\partial X}{
 \partial F_{\mu\lambda}} \nabla^\nu F_{\mu\lambda}
 + \frac{\partial X}{\partial \phi} \partial^\nu \phi
 \right] - F_{\ \lambda}^{\nu} \nabla_\mu
 G^{\mu\lambda}
 - G^{\lambda\mu} \nabla_\mu F_\lambda^{\ \nu} \nn\\
 &=& - g^{\nu\sigma} G^{\mu\lambda} \left( \frac12
 \nabla_{\sigma} F_{\mu\lambda} + \frac12 \nabla_\mu
 F_{\lambda\sigma} + \frac12 \nabla_\lambda F_{\sigma\mu}
 \right) - F_{\ \lambda}^{\nu} J^\lambda \\
 && \qquad\qquad\qquad\qquad\qquad\qquad\qquad\qquad
 + \nabla_\mu \left[ u^\mu u^\nu ( \rho_m + p )
 + p \, g^{\mu\nu} \right] \nn\\
 &=& - F_{\ \lambda}^{\nu} J^\lambda + \nabla_\mu
 \left[ u^\mu u^\nu ( \rho_m + p )+ p \, g^{\mu\nu}
 \right] = 0 \,, \nn
\eea
in which the fluid 4-velocity is $u_\mu = \delta^t_\mu \, \sqrt{- g_{tt}}$, and the right-hand-side has been simplified using the other field equations.

To solve these we adopt the following {\em ans\"atze} for the Maxwell field,
\be
 G^{tr} = \frac{D(r)}{ L^2 \sqrt{-g_{tt}g_{rr}}} \,,
\ee
and the metric,
\bea
 \exd s^2 &=& -e^{2a(r)} \exd t^2 + \frac{L^2 \exd r^2}{
 1 +r^2 - {\kappa^2m(r)}/{4\pi Lr}}
 + L^2 r^2 \exd\Omega^2 \,.
\eea
The equations for the unknown functions $D(r)$, $a(r)$, $m(r)$, $\rho_m(r)$, $\rho_c(r)$, $p(r)$ and $\phi(r)$ then simplify to a set of coupled ordinary differential equations. The Maxwell equation becomes
\begin{equation}
 D' + \frac{2D}{r} - \frac{L^3\rho_c}{ \sqrt{1 +r^2 -
 {\kappa^2m}/{4\pi rL}}} = 0 \,,
\end{equation}
and the dilaton equation is
\begin{equation}
 \left(1 + r^2 -\frac{\kappa^2m}{4\pi Lr} \right)
 \left(\phi'a' + \frac{2\phi'}{r} + \phi''\right)
 + \phi'\left( r -\frac{\kappa^2m'}{8\pi Lr}
 + \frac{\kappa^2m}{8\pi Lr^2} \right)
 + \frac{\mathcal{\hat T}(X^2-1)}{X} = 0 \,,
\end{equation}
while the Einstein equations are
\bea
 m' - 4\pi\rho_m  L^3r^2 - \frac{4\pi L\mathcal{\hat T}}{X} \, (1-X) - L \pi \left( 1- \frac{\kappa^2m}{4\pi Lr} + r^2
 \right)(\phi')^2 &=& 0 \\
 \hbox{and} \quad
 2a' - \frac r4 \, (\phi')^2 - \frac{ \frac{\kappa^2
 m}{4\pi L} - 2r^3 + r^3 \left( L^2 \kappa^2 p
 - \frac{\mathcal{\hat T}}{X} \, (1-X) \right) }{ 2r^2
 \left( 1 + r^2-  {\kappa^2m}/{4\pi Lr}  \right)}
 &=& 0 \,.
\eea
Finally, conservation of energy-momentum becomes
\be
 p' + (p + \rho_m ) \, a' - \frac{\rho_c \, e^{\phi/2}
 \sqrt{1-X^2}}{ \kappa^2 L^2 \sqrt{ \left( 1 +r^2  -
 {\kappa^2m}/{4\pi Lr} \right)}} = 0 \,,
\ee
and the function $X$ reduces to
\be
 X^{-2} = {1+\frac{D^2e^\phi}{\mathcal{ \hat T}^2L^4}} \,.
\ee

\FIGURE[!t]{
 \centering
 \includegraphics[width=70mm]{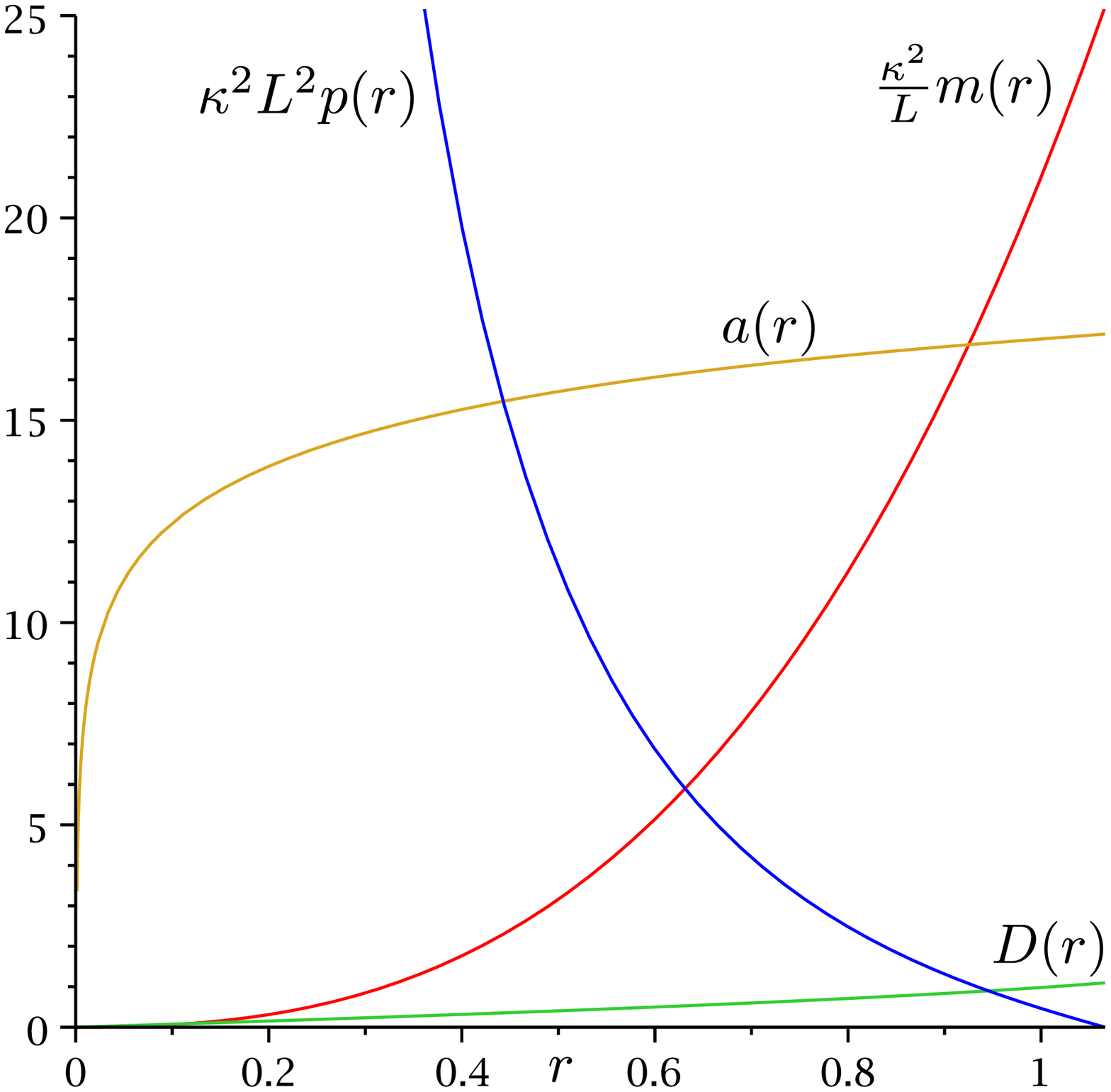}
 \caption{A plot of the properties of the interior of an incompressible star for $Q=1.25, \mathcal{\hat T}=1/1000$. The assumption of constant mass and charge density is used as an illustrative example, since we are not interested in the exact values of stellar masses or radii (colour online).\label{fig:starfunctions}}
}

These equations are to be integrated subject to a choice of equation of state, $p = p (\rho_c)$ and $\rho_m = \rho_m(\rho_c)$, and in practice we solve these numerically (see however \cite{MH} for a semi-analytic approach to a similar problem). Figure \ref{fig:starfunctions} plots a solution assuming an incompressible equation of state, with $\rho_m$ and $\rho_c$ constants. For pure gravity and asymptotically flat spacetimes the mass at which a star forms a black hole with this equation of state is the largest possible \cite{buchdahl}, and so for simplicity we use it here for all explicit numerical integrations in the hopes that it also provides the most massive possible stars in this more complicated theory. However we do not believe this plays a crucial role for the purposes of identifying how the solutions scale with changes to $L$.

The boundary conditions to be satisfied require all fields be regular at $r = 0$, and so $a(0) = a'(0) = m(0) = m'(0) = \phi'(0) = D(0) = 0$. They must also be continuous across the stellar surface --- which is defined as the radius, $r = r_s$, where $p(r_s) = 0$. For instance, for the Maxwell field the function $D(r)$ must be matched to the exterior solution, for which $D_{\rm ext}(r) = Q/r^2$, and so $D(r_s) = Q/r_s^2$. The metric functions similarly continuously match to the exterior solutions described in the previous sections, and the stellar mass, $M$, is identified as the ADM mass for the asymptotic external geometry.

All told, for a given equation of state we have a three-parameter family of initial conditions, corresponding to our choice for the central values, $\phi(0)$ and $p(0)$, as well as the periodicity, $\beta'$, of the euclidean time direction at infinity. (There are four parameters if we also include the local charge-to-mass ratio, rather than thinking of $\rho_c(0)/\rho_m(0)$ as an equation of state.)  Integrating the equations then allows all other parameters to be computed for $r > 0$. For AdS/CFT applications these three parameters are instead regarded as all being specified at infinity, by giving $\beta$, $\phi_0$ and the stellar mass, and then integrating the solutions towards smaller $r$. ($Q$ is the fourth parameter if $\rho_c(0)/\rho_m(0)$ is also regarded as to be specified.) Since we build the star from charged matter, it is physically clear that no stable solution should be possible unless the mass is large enough compared with the charge to allow gravitational attraction to overwhelm electrostatic
repulsion.

If we do scale the system size, $L\rightarrow\lambda L$, then dimensional analysis shows that the field equations remain invariant provided we also rescale, $\kappa^2p\rightarrow\lambda^{-2}\kappa^2p$, $\kappa^2\rho_m\rightarrow\lambda^{-2}\kappa^2\rho_m$, $\kappa^2m\rightarrow\lambda\kappa^2m$, and $\rho_c\rightarrow\lambda^{-3}\rho_c$.  This implies that calculable quantities like the stellar size, $r_s$, and stellar mass, $M$, must obey scaling relations like
\begin{equation}
 \kappa^2 M = L \, \cF\left( L^2 \kappa^2 p(0), L^3 \rho_c(0), \mathcal{\hat T}, \phi_0 \right) \,,
\end{equation}
for some dimensionless function $\cF$.

\subsection{Energetics of the transition} \label{sec-phasediagram}

In this section we work out the free energy of the black hole and stellar regimes and identify when thermodynamics prefers the crossover to be from one phase to the other.

\subsubsection*{Comparison of the free energies}

The free energy in both phases is computed in appendix \ref{sec-fecalcs}, leading to the following expressions
\bea
 F_\ssT(T) &=&  \frac{4\pi LT \beta}{\kappa^2}
 \int_{r_h}^{r_\infty} \exd r \; e^{\xi(r)/2} r^2 \left[
 3 + \frac{\mathcal{\hat T} (X-1)}{X}\right] \\
 F_\ssS(T) &=& \frac{4\pi LT \beta' }{\kappa^2} \left\{\int_{0}^{r_s} \frac{ \exd r e^{a(r)/2} r^2}{
 \sqrt{1 + r^2 - {\kappa^2m}/{4\pi Lr}}} \right. \left[ 3 + \frac{ \mathcal{\hat T} (X-1)}{X} +
 \frac{\kappa^2L^2}2 ( p - \rho_m ) \right] \nn\\
 &&\qquad\qquad\qquad\qquad\qquad\qquad
 + \left. \int_{r_s}^{r_\infty} \exd r \, e^{\xi(r)/2} r^2 \left[ 3 + \frac{\mathcal{\hat T} (X-1)}{X} \right] \vphantom{ \frac{\exd r e^{a(r)/2} r^2}{\sqrt{1 - \frac{\kappa^2m}{4\pi Lr} + r^2}}} \right\} \,,
\eea
where $F_\ssT$ (or $F_\ssS$) is the free energy computed with the black hole (or stellar) solution. $\beta'$ is the periodicity of the time circumference for the stellar geometry, in the same way that $\beta$ is for the black hole.

Both solutions are labeled by their total mass and charge, the asymptotic value, $\phi_0$, for the dilaton, and by the periodicity, $\beta$, of their time direction (in euclidean signature). As discussed earlier, $SL(2,R)$ invariance guarantees $Q$ and $\phi_0$ only appear through the combination $Q^2 \, e^{\phi_0}$, in principle leaving three independent parameters.

For the black-hole solution the total mass can be traded for the horizon radius, $r_h$, which should be regarded as a function of the other externally fixed variables. Regularity of the (euclidean signature) geometry at $r = r_h$ then also gives $\beta$ as a function of these other quantities.

For a stellar solution, rather than specifying quantities like the central pressure deep within the star and integrating out to larger $r$, we instead regard the asymptotic mass, charge and dilaton field to be the quantities specified by the CFT parameters, and integrate in towards smaller values of $r$ to find the properties interior to the star. In principle the total stellar mass can be traded for the stellar radius, $r_s$, given a specific internal equation of state.

Since both $r_s$ and $r_h$ can be computed for a given set of asymptotically specified parameters, they can be compared with one another. Physically we expect $r_h < r_s$, since otherwise the metric function, $g_{tt}$, vanishes before the stellar surface is reached, making the solution a black hole. The left panel of figure \ref{fig:fediffcrittemp} verifies this expectation in a comparison between $r_s$ and $r_h$ for shared external parameters, in the particular case where the geometry external to the star and black hole has the $z = 5$ attractor form given above. The calculation is much easier in this case because the quantity $X$ is then a constant, and $r_s$ for the stellar solution can be computed from the condition that $X(r = r_s) = X_h$, with $r_h$ computed from $h(r_h) = 0$ in this geometry. For small $r$ we do this by solving for $r_h$ numerically using the implicit expression for $h(r)$ in \eqref{smallrhansatz}.

\FIGURE[ht!]{
 \centering
 \includegraphics[width=60mm]{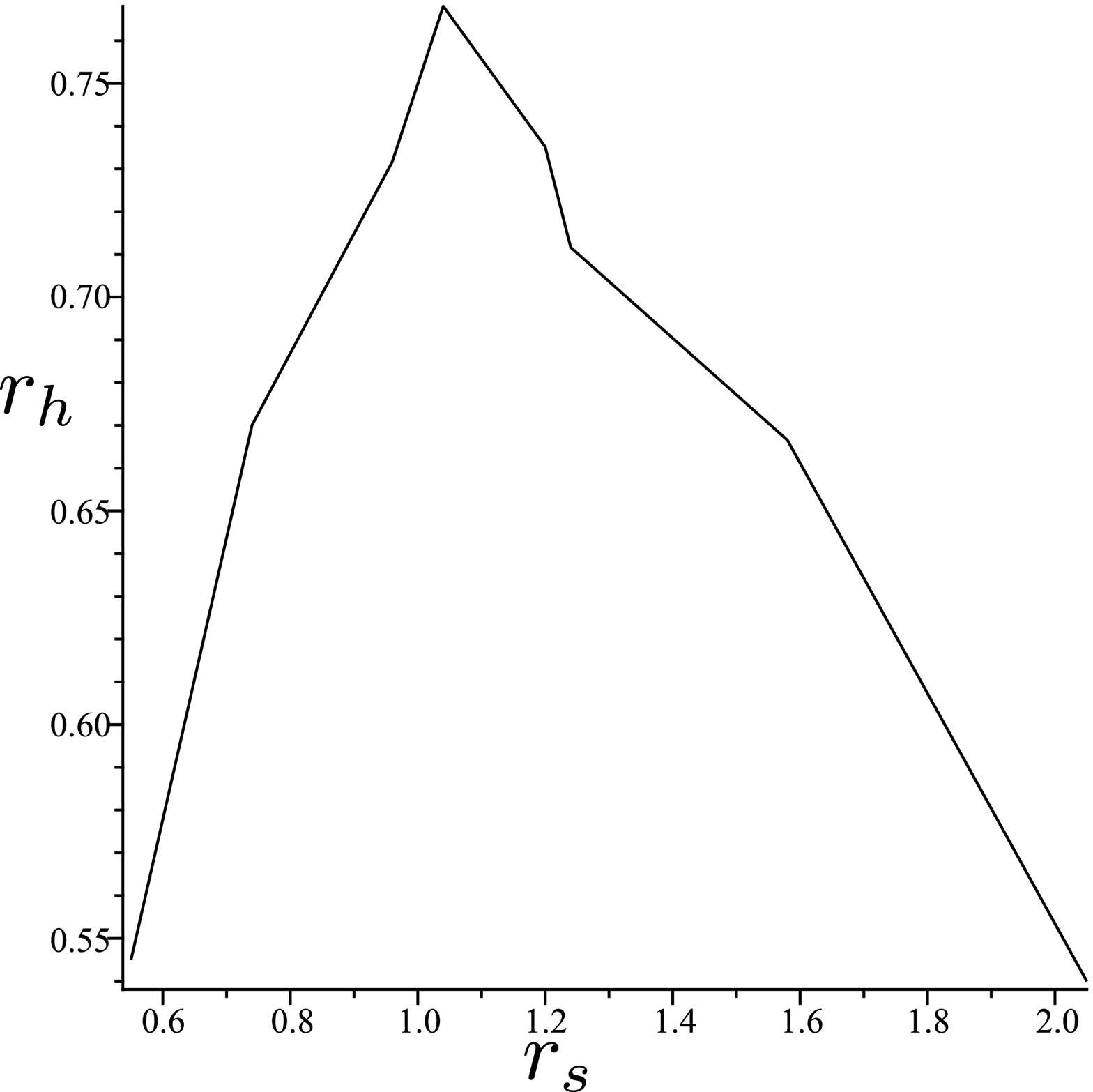}
 \includegraphics[width=70mm]{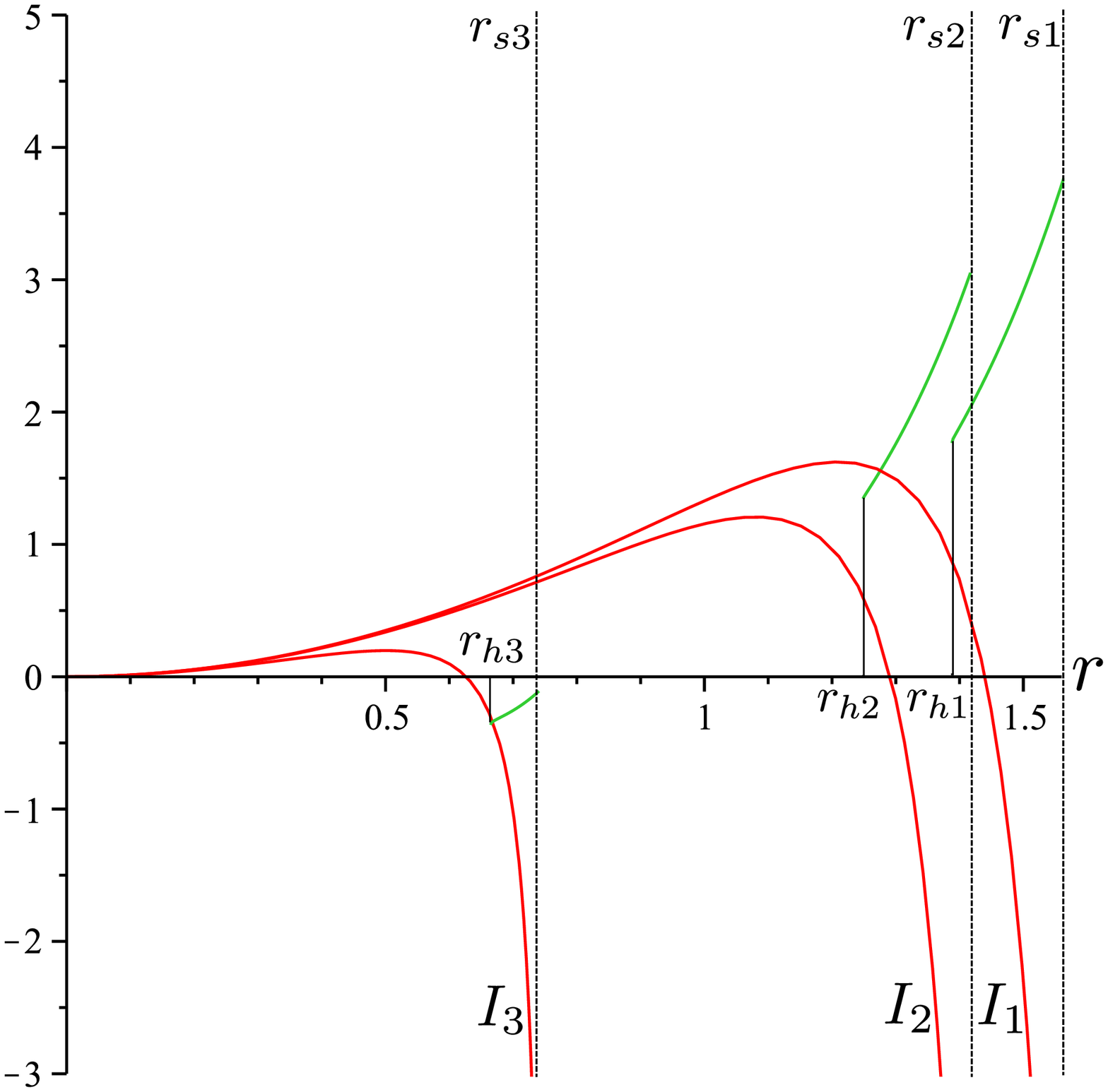}
 \caption{Left panel: A comparison of the black-hole horizon radius, $r_h$, and stellar size, $r_s$, for a shared choice of asymptotic boundary conditions in the special case that the external geometry just outside the star or black hole has the $z=5$ asymptotic form (see text for discussion).
 Right panel: The integrands for the two terms in \eqref{fediff} for the free-energy difference between stellar and black-hole geometries. The red curves labeled
 $I_1, I_2,$ and $I_3$ give the contribution of the stellar interior, for the following values of central pressure: $\kappa^2L^2p(0)= 10^5, 10,$ and $1$ for which $r_s=1.56, 1.42,$ and $0.74$ respectively. The remaining three (green) curves show the contribution from the exterior geometry. These curves rise sharply at $r_h$, which can be seen to occur at $r_h= 1.39, 1.25,$ and $0.66$ respectively. All curves stop at the value $r=r_s$ appropriate to the solution in question. The free energy difference evaluates to $\Delta F_1=-.21$, $\Delta F_2 = -.01$, and $\Delta F_3=.045$.  Left panel uses $\mathcal{\hat T}=10^{-5}$, and right panel uses $\mathcal{\hat T}=10^2$ (colour online). \label{fig:fediffcrittemp}}.
}

Because we specify the same quantities at large distances and integrate into the interior for both stellar and black hole solutions, both the black hole and stellar phase will share the same exterior geometry for $r > r_s$ at fixed $Q, \mathcal{\hat T}$ and $\phi_0$. Because of this $\beta' = \beta = 1/T$, although for the black-hole solution $\beta$ is not regarded as being independent of the other quantities.

Given these observations, appendix \ref{sec-fecalcs} shows that the free-energy difference, $\Delta F := F_\ssT - F_\ssS$, takes the form
\bea
 \Delta F &=& \frac{4\pi L}{\kappa^2}
 \left\{ - \int_{0}^{r_s} \exd r\; \frac{ e^{a(r)/2} r^2}{
 \sqrt{1 + r^2 - {\kappa^2m}/{4\pi Lr}}}
 \right. \left[ 3 + \frac{\mathcal{\hat T}(X-1)}{X}
 + \frac{\kappa^2L^2}2 (p - \rho_m ) \right] \nn\\
 &&\qquad\qquad\qquad\qquad\qquad\qquad
 + \left. \int_{r_h}^{r_s} \exd r \; e^{\xi(r)/2} r^2 \left[ 3 + \frac{\mathcal{\hat T}(X-1)}{X} \right]
 \vphantom{\frac{\exd r e^{a(r)/2} r^2}{ \sqrt{1 + r^2 - {\kappa^2m}/{4\pi Lr}}}} \right\} \,. \label{fediff}
\eea
The right panel of figure \ref{fig:fediffcrittemp} plots the integrands of this expression, with the red curves labeled $I_1$, $I_2$ and $I_3$ giving the contribution of the stellar interior ($0 < r < r_s$) for successively smaller choices of central pressure. This shows that this integral contributes a negative contribution to $\Delta F$ for large central pressures (because of the explicit negative sign in front of the integral in \pref{fediff}), which becomes less negative for smaller central pressures. The three smaller green curves plot the contribution of the external geometry $(r_h < r < r_s)$ for the same external parameters, showing these contribute a decreasing (and eventually negative) amount to $\Delta F$, that decreases with decreasing central pressure.

The figure shows that it is the black-hole phase that has the lower free energy (negative $\Delta F$) for large enough central pressures, although the free-energy differential between the two solutions falls with falling stellar mass (and so also central pressure). Because the black-hole free energy has a minimum temperature and the stellar free energy does not, the free energy difference eventually changes sign, leading to an eventual crossover to the stellar phase (positive $\Delta F$) at lower central pressures.

The physical picture that emerges is as follows. Imagine we fix the system size, $L$, and start the system at high temperatures with parameters chosen so that a $z=5$ region exists for sufficiently small $r$. For sufficiently large temperatures only a black hole solution is possible, because the euclidean time direction at large $r$ has a small enough circumference to ensure that integrating the field equations down to smaller $r$ leads to a zero of $g_{tt}$ for some $r_h$ that is larger than the would-be size $r_s$ of the star having the same mass and charge. Alternatively, large temperature means large $r_h$ and for very large $r_h$ no stellar solutions are possible because large $r_h$ corresponds to a total mass above the Chandrashekar limit, for which there is no central pressure large enough to support a star.

As we bring down the temperature, the asymptotic circumference in the $t$ direction increases, and so the value of $r_h$ where $g_{tt}$ vanishes decreases. Eventually we reach a point where $r_h$ lies sufficiently below $r_s$ to ensure that the total mass is below the Chandrashekar limit, in which case the stellar phase is permitted. Since the central pressure is high, the stellar phase at first has a higher free energy than the black-hole solution, and so the system remains in the black-hole phase. As $T$ continues to fall, $\Delta F$ becomes less negative until eventually the stellar phase becomes preferable. Precisely where this occurs likely depends on the details of the stellar equation of state chosen (although we argue below that this is not important for the purposes of identifying how the transition temperature depends on system size). At sufficiently small temperatures no black hole solution exists at all as an alternative to the stellar solution.

\subsection{Shape of the transition curve}
\label{sec:curveshape}

We now seek the shape of the transition curve, $T_s(L)$, that is defined by the condition $\Delta F = 0$. Notice that the free-energy expressions expose the variables on which $\Delta F$ depends,
\begin{equation}
 \Delta F = \frac{4\pi L }{\kappa^2}
 \Bigl[ \cF_{\ssT}(r_h, Q \, e^{\phi_h/2}, \mathcal{\hat T})
 - \cF_{\ssS}(r_h, Q \, e^{\phi_h/2}, \mathcal{\hat T}) \Bigr] \,.
\end{equation}
and we do not separately list $r_s$ because, as we have seen, $r_s$ and $r_h$ are not independent of one another since they are both proxies for the total stellar (or black hole) mass.

If, however, the geometry of the region $r_h < r < r_s$ lies within the $z=5$ attractor regime, as is required for successful description of quantum Hall temperature scaling, even fewer parameters turn out to be independent. This is because $X = X(Q \, e^{\phi/2}, \mathcal{\hat T})$ evaluates to an $r$-independent constant in this regime, whose value $X(r) = X(r_h) := X_h$ is completely determined by $\mathcal{\hat T}$. $X$ is kept constant in this near-horizon regime because the dilaton adjusts itself as $Q$ is varied to keep $Q \, e^{\phi/2}$ fixed at any given $r$, allowing any dependence on $Q \, e^{\phi_h/2}$ to be traded for a dependence on $r_h$ and $\mathcal{\hat T}$, and so
\begin{equation}
 \Delta F = \frac{4\pi L }{\kappa^2}
 \Bigl[ \cF_{\ssT}(r_h, \mathcal{\hat T})
 - \cF_{\ssS}(r_h, \mathcal{\hat T}) \Bigr] \,. \label{tensiondeptrans}
\end{equation}

Finally, we know that $\mathcal{\hat T}$ drops out of the field equations and the integrand of the free energy when it is either very large or small. So long as this is true, we are effectively left with the free energy difference
\begin{equation}
 \Delta F
 = \frac{4\pi L}{\kappa^2} \Bigl[\cF_{\ssT}(r_h) -
   \cF_{\ssS}(r_h) \Bigr] \,. \label{onlyhorizondeptrans}
\end{equation}
This shows that the condition $\Delta F = 0$ may be regarded as a condition that $r_h$ takes on a fixed value (as anticipated in earlier sections).

Now the discussion follows the discussion below eq.~\pref{z5temp}, which tells us we may trade a dependence on $r_h$ for a dependence on $LT$ and $r_c$, with $r_c$ depending differently on $L$ depending on whether or not $\mathcal{\hat T}$ is large or small, so the free energy difference becomes
\begin{equation}
 \Delta F =
 \frac{4\pi L}{\kappa^2} \Bigl[ \hat\cF_{\ssT}(LT,r_c) -
 \hat\cF_{\ssS}(LT,r_c) \Bigr] \,.
\end{equation}
Recalling equations \eqref{z5Tcrossoverdep}, when $\mathcal {\hat T} \gg 1$, the $L$-dependence of $r_c$ drops out and if there exists a $T_s$ for which $\Delta F(LT_s,r_c) = 0$, then clearly $T_s \propto 1/L$ is a simple consequence.  Furthermore, if we assume that this vanishing free energy difference occurs when the solutions in section \ref{z5soln} are valid --- figure \ref{fig:h1omegaTTprhdep}, for $\omega(r_h)\approx 4$ gives a good indication where this solution is valid --- then trading the horizon radius $r_h$ for $LT$ and $r_c$ makes the difference take the form
\begin{equation}
 \Delta F =
 \frac{4\pi L}{\kappa^2} \Bigl[ \hat\cF_{\ssT}(LTr_c^4) -
 \hat\cF_{\ssS}(LTr_c^4) \Bigr] \,.
\end{equation}
If $\mathcal{\hat T} \ll 1$ in this case, then a $T_s$ for which $\Delta F(LT_sr_c^4) = 0$, then $T_s\propto 1/L^5$.

Notice that this argument is fairly robust, since it rests on only a small number of assumptions.  The first is that the near-horizon geometry is well-described by the $z=5$ attractor geometry, since then the attractor mechanism guarantees the value of $X$ is independent of the boundary value of $\phi_0$ and $Q$. This assumption is always satisfied for the parameter regime describing quantum Hall systems, since it is this regime that ensures the success of the prediction $p = 2/z$ for the scaling exponent.

The second, more model-dependent, assumption is that the dimensionless tension, $\mathcal{\hat T} = \kappa^2 L^2 \cT$ is either very large or very small. (If $\mathcal{\hat T}$ is large, then the discussion below eq.~\pref{feexpansion2} shows that we must also require that $Q^2 e^{\phi_h} X_h/\mathcal{\hat T} \ll 1$.) The robustness of these choices shows that it should be generic that $T_s \propto L^{-1}$ for large dimensionless tension. The scaling $T_s \propto L^{-5}$ occurs when the dimensionless tension is very small and the near-horizon geometry is well-described by the $z=5$ attractor geometry.  Finally, it should also be possible to choose special values $\mathcal{\hat T} \sim 1$ for which these behaviours fail.

These choices fall within the domain of validity of the calculation, which was defined by two separate conditions. The first is for small $\mathcal{\hat T}$, in which $Q^2 e^{\phi_h}$ is fixed by the attractor to a number independent of $\mathcal{\hat T}$, and the free energy has a transition that is only dependent on the quantity $LT$.  The second is for large $\mathcal{\hat T}$, for which $Q^2 e^{\phi_h}$ is again fixed by the attractor, although this time proportional to $\mathcal{\hat T}$ because of \eqref{hightenattractedvalue}.  In both these cases $Q$ can be large if $e^{\phi_h}$ is small enough since $SL(2,R)$ ensures they always appear together.

\subsubsection*{Conductivity in the low-temperature regime} \label{sec-comparison}

It is one thing to have a new low-temperature regime, but does it have the right properties to describe real quantum Hall systems? Answering this is a research project in itself, but we suffice here to argue that conductivities become independent of temperature in this regime, as they must to agree with observations.

To see this, consider the formula for the Ohmic conductivity given in appendix \ref{sec-condcalc},
\be
 \sigma_{\theta\theta}^2 \simeq \sin^2\theta e^{-2 \tilde\phi}\mathcal T^2 \ell^8 + \frac{e^{-\tilde\phi_0}Q^2\ell^4}{L^4r_h^4} \,.
\ee
For the black-hole solution quantities like $r_h$ can be traded for temperature, because the nonsingularity of the horizon geometry relates the periodicity, $\beta$, to other geometrical quantities. The important observation is that this is no longer true for the stellar solutions, because there is no horizon on which to be singular. As a result, none of the variables appearing above depend on temperature, and so both Hall and Ohmic conductivities should be temperature-independent.

\subsubsection*{Thermodynamic signature}

A noteworthy feature of the above description is that the AdS/CFT picture predicts that derivatives of $F$ are discontinuous across the transition to a temperature-independent regime. This makes this a first-order phase transition, with definite implications for the thermodynamic properties of the quantum Hall fluid.

This discontinuity is seen in figure \pref{fig:federivative}, which plots the temperature derivative
\bea
 \frac{\exd F_T(T)}{\exd T} &=& \frac{\exd F_T(T)}{\exd r_h}\frac{\exd r_h}{\exd T} \nn\\
 &=& -\frac{4\pi L^2}{\kappa^2}\frac{e^{\xi(r_h)/2}r_h^2}{(e^{\xi(r_h)/2}h_1(r_h))'}\left[3+\mathcal{\hat T}\frac{X_h-1}{X_h}\right],
\eea
obtained by implicitly differentiating the black-hole expression, eq.~\eqref{BHfreeenergy}. The second line of this equation uses eqs.~\eqref{smallrhtemp} and \eqref{BHfreeenergy}. Here primes denote derivatives with respect to $r_h$. Notice in particular the kink as we approach the minimum temperature, caused by the vanishing of $\exd T/\exd r_h$ at $T = T_\star$. There is no similar near-divergence in the stellar-phase free-energy derivative, indicating that it is the black hole free energy that dominates the discontinuity of $\exd F/\exd T$ near the transition (see, {\em e.g.} figure \ref{fig:fediffcrittemp}).

Although it might seem odd to have a discontinuous transition in a finite-sized system, such as this, the discontinuity is a consequence of the large-$N$ limit that is implicit on the CFT side of the AdS/CFT correspondence when working at the classical level on the gravity side.

\FIGURE[h]{
 \centering
 \includegraphics[width=60mm]{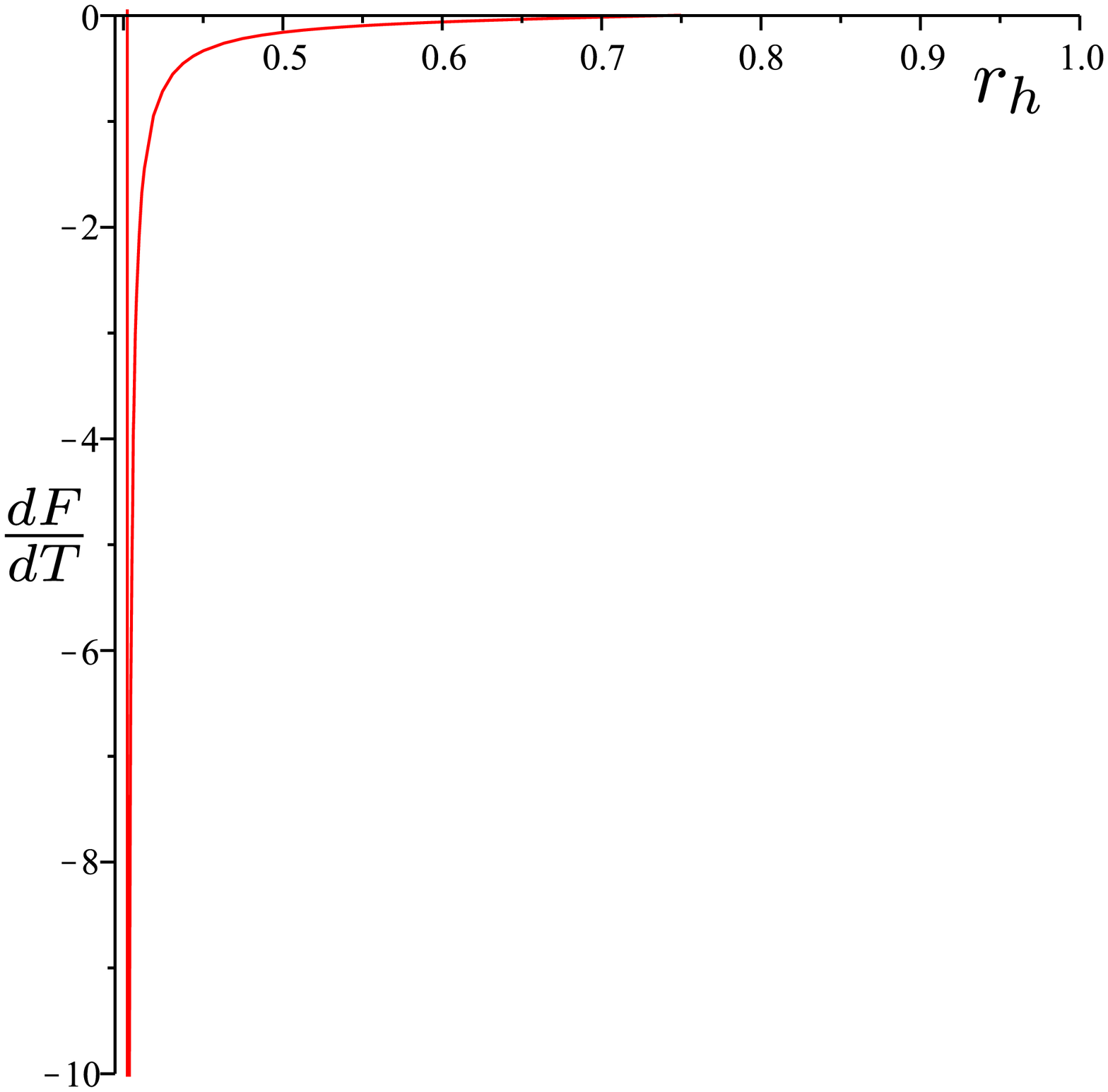}
 \caption{Derivative of the free energy in the black hole phase as a function of horizon position (in units of ${4\pi L}/{\kappa^2}$). Notice its singular behaviour as the temperature approaches the minimum of the black hole phase. \label{fig:federivative}}.
}

\section{Conclusions} \label{sec-conclusion}

We present here a method for modeling finite-size effects within the AdS/CFT description proposed for quantum Hall systems in ref.~\cite{Hallography}. We do so following standard tools \cite{Witten-gauge}, but adapted to CFTs with nonzero chemical potential. We do so in order to see whether the very successful AdS/CFT description \cite{QHEfinitesizeexp}, $p = 2/z = 0.4$, of a scaling exponent for how a critical Hall resistance scales with temperature, is consistent with the existence --- in the same experiments that measure $p$ --- of a crossover to a low-temperature regime for which the Hall resistance is temperature independent. The transition temperature, $T_s$, between these regimes is measured to depend on system size, $L$, according to $T_s \propto 1/L$, and we ask whether this is consistent with the holographic identification of $z = 5$ as a dynamical scaling exponent (which would naively predict $T_s \propto 1/L^5$).

We find the holographic AdS/QHE picture is consistent with a transition to a $T$-independent Hall conductivity at very low temperatures. Remarkably, the transition is described on the gravity side by the transition from a black hole configuration to a description in terms of an asymptotically AdS charged star, extending similar treatments of stellar objects in AdS/CFT to finite-size systems \cite{Arsiwalla:2010bt,Hartnoll:2010gu}.

Furthermore, we find that the phenomenology of the transition to finite-size effects --- and in particular the size-dependence of the transition temperature, $T_s \propto 1/L$ --- {\em can be} consistent with the prediction $p = 2/z$ and the use of $z = 5$, if the dimensionless brane tension, $\mathcal{\hat T} = \kappa^2 L^2 \cT$, appearing on the AdS side is sufficiently large. When this tension is negligible, naive scaling prevails and the transition temperature satisfies $T_s \propto 1/L^5$, but when $\cT$ is large enough its presence alters naive scaling relations in such a way as to instead predict $T_s \propto 1/L$.

Tension changes the scaling relations by modifying the $L$-dependence of the radius, $r_c$, that defines the crossover between the $z = 5$ near-horizon geometry (relevant to the IR limit) to the asymptotic $z = 1$ geometry that obtains at large $r$ (and is relevant in the UV). Recalling that the existence of different UV and IR scalings relies on the black hole being charged, it is no surprise that $r_c$ grows with black-hole charge: $r_c^4 \propto Q^2 = (\rho_\CFT L^2)^2$, which implies {\em in the absence of other scales} that $r_c \propto L$ when the CFT charge density $\rho_\CFT$ is held fixed. A sizeable tension instead changes this relation to the $L$-independent result: $r_c^4 \propto Q^2/\mathcal{\hat T} = (\rho_\CFT/\kappa^2 \cT)^2$.

The $L$-dependence of $r_c$ makes a difference because within the near-horizon $z=5$ geometry the relation between the temperature and horizon radius, $r_h$, is $T \propto (r_h^5/r_c^4 L)$. For fixed $r_h$ this predicts $T \propto 1/L^5$ in the naive scaling limit where $r_c \propto L$. But it instead predicts $T \propto 1/L$ when $r_c$ is $L$-independent.

Notice that because $r_c$ does not depend on $T$ none of these considerations about $r_c$ modify the Hall conductivity's temperature dependence, which traces its roots to the relation $\sigma_{xy} \propto 1/r_h^2$ and $T \propto r_h^5$. It {\em does} modify how temperature scales with $L$, however, for any quantity (like the transition temperature between low- and high-temperature phases) that is defined by a fixed value for $r_h$.

In the AdS/QHE description the transition is predicted to be a first-order phase transition, for which derivatives like $\exd F/\exd T$ are discontinuous.\footnote{Discontinuities are allowed at finite volume in holographic descriptions due to the large-$N$ limit that is implicit, on the CFT side, when using semiclassical reasoning (as we do) on the AdS side.} This has implications for the thermodynamics of the quantum Hall fluid at the transition regime, whose presence tests the entire framework. Although such effects would be experimentally challenging to find, their detection would be worth the effort.

Likewise, since both $T_s \propto 1/L$ and $T_s\sim L^{-5}$ are possible, depending on the size of $\mathcal{\hat T}$, it would be worth better understanding what the parameter $\cT$ captures on the CFT side, in order to suggest how to obtain samples for which $T$ scales differently than $1/L$. Our difficulty in doing so at present is a limitation of our phenomenological approach within which the AdS field equations and brane properties are assumed without reference to a UV completion (which would entail a full embedding into string theory).

This is one aspect of an important missing step in the AdS/CFT description of quantum Hall systems: a precise enunciation of its bounds of validity. If this were known we would also know what conditions were sufficient for the inter-plateaux behaviour to be universal and not sample-dependent. Ref.~\cite{QHEfinitesizeexp} sheds some experimental light on this issue, since it also shows how scaling changes as the samples are doped. Whenever scaling with $T$ is robust --- {\em i.e.} scaling lasts over two decades of temperature --- its power is given by the universal value $0.42 \pm 0.01$. But such robust scaling is only found for a relatively small range of doping; a range for which scattering from the doped atoms is likely to dominate the conductivity \cite{QHEfinitesizeexp, Alloy}. For other dopings the temperature-dependence of the conductivity is more sample-dependent. It is important to understand from a microscopic point of view what it is about the quantity of dopant that promotes a universal 
description.

But the larger point is that the holographic AdS/QHE model presented in \cite{Hallography} appears to agree well with more observations than those that are guaranteed by construction through its incorporation of emergent duality symmetries. We believe that the experimental success of these predictions for $p$ and $T_s(L)$ provide strong evidence that holographic models provide the natural theoretical language for describing quantum Hall systems. As such, quantum Hall systems are also likely to be a rich environment for testing AdS/CFT methods.

Many steps in the AdS/QHE program remain incomplete. Perhaps most important is a robust treatment of disorder in holography, since this plays a central role in producing DC conductivity. Our present tools for exploring holographic conductors remain strongly hampered by an inability to incorporate disorder in a simple way. Work on all these issues proceeds apace.

\section*{Acknowledgements}

We thank Gabor Cs\'athy, Brian Dolan, E. Fradkin, Michael Hilke, Janet Hung, S. Kivelson, Sung-Sik Lee, Rob Myers and Yanwen Shang for useful conversations. This research has been supported in part by funds from the Natural Sciences and Engineering Research Council (NSERC) of Canada. Research at the Perimeter Institute is supported in part by the Government of Canada through NSERC and by the Province of Ontario through the Ministry of Research and Information (MRI).

\appendix

\section{Conductivity for finite size}
\label{sec-condcalc}

Since fixed-$r$ surfaces in the finite-size case are spheres rather than planes, linear-response theory is trickier since it is not possible to turn on a constant perturbing electric field. This section discusses how we think about performing this calculation.

To calculate the conductivity at finite size we make two simplifying assumptions.
\begin{itemize}
\item We use the $z=5$ attractor solution of ref.~\cite{Hallography}, which involves two separate choices. The first choice is that $r_h$ is large, so that the black-brane solution is a good approximation to the black-hole solution. The second is to use the numerical value $z = 5$, that ref.~\cite{Hallography} shows is the near-horizon solution for dilaton-gravity coupled to the DBI action for a non-probe brane. (We do so for the motivations given in the main text).
\item The second assumption is to focus the conductivity calculation on a small patch of the sphere in the region of the equator. We do this to avoid any singularities ({\em e.g.} near the poles) that inevitably arise if a global electric field is applied everywhere on a sphere.
\end{itemize}
We assume the finite size background to be
\begin{equation}
 \exd s^2 = -h_0 \frac{r^{10}}{r_c^8} \left( 1 - \frac{r_h^7}{r^7}
 \right) \exd t^2 + \frac{L^2 \exd r^2}{h_0 r^2 \left( 1- \frac{r_h^7}{r^7} \right)} + r^2 L^2 \left( \exd\theta^2 + \sin^2\theta \exd\varphi^2 \right) \,, \label{dimlesscoords}
\end{equation}
which has the temperature
\begin{align}
 4\pi T=\frac{7h_0r_h^5}{r_c^4L}.
\end{align}

Again, this solution is valid when $r_h\gg1$ and $r_h^4\ll\frac{Q^2e^{\phi_0}}{\mathcal{\hat T}^2}$ (which is not the same dilaton in what follows, since we use the probe brane approximation by taking a stack of branes to generate the background, and a seperate brane to probe the geometry), that is away from the minimum temperature (and finite size phase transition.)

To get the conductivity, we place a flux through the surface of the sphere from pole to pole.  That is, along the $\theta$ direction.  Corresponding to
\begin{align}
 J^\theta&=\sqrt{-g}G^{r\theta} \nn\\
 E_\theta(\theta)&=F_{t\theta}\\
 Q&=\sqrt{-g}G^{rt} \,,\nn
\end{align}
where $E_\theta(\theta)$ has some angular dependence that satisfies its Maxwell equation.  Proceeding with the identical arguments of \cite{Hallography} we get the conductivity
\bea
 \sigma_{\theta\theta}^2 &=& \left[ \mathcal T^2\ell^8e^{-2\tilde\phi} \frac{g_{\varphi\varphi}}{g_{\theta\theta}} +
 \frac{e^{-\tilde\phi}Q^2\ell^4}{g_{\theta\theta}^2}
 \right]_{r = r_h}\nn\\
 &=&\sin^2\theta e^{-2 \tilde\phi_0}\mathcal T^2 \ell^8+\frac{e^{-\tilde\phi_0}Q^2\ell^4}{L^4r_h^4}.
\eea
For low enough temperatures the second term is dominant, provided
\bea
 e^{-2\tilde\phi_0}\mathcal T^2 \ell^8&\ll&\frac{e^{-\tilde\phi_0}Q^2\ell^4}{L^4r_h^4} \nn\\
 \hbox{or, equivalently} \quad e^{-2\phi_0}\mathcal T^2 \kappa^8&\ll&\frac{e^{-\phi_0}Q^2\kappa^4}{L^4r_h^4}\\
 \hbox{and so} \quad
 r_h^4&\ll\frac{Q^2e^{\phi_0}}{\mathcal{\hat T}^2},\nn
\eea
where the second line uses $\tilde\phi=\phi + 4\log(\ell/\kappa)$ and we set $\sin\theta=1$ since we are calculating the conductivity near the equator.  This is the exact same condition as to be in the $z=5$ region from \eqref{z1approx}.
%
%
Expressed as a function of temperature (for the black hole geometry) the conductivity in the finite size case using \eqref{z5temp}, is then
\bea
 \sigma_{\theta\theta} &=&\kappa^2\rho_{\CFT} \, e^{-\phi_0/2}\times\begin{cases} \left[\frac{\kappa^4\mathcal T^2}{(7h_0)4\pi LT\rho_\CFT^2e^{\phi_0}}\right]^{2/5} & \text{if $\mathcal{\hat T}\gg 1$,}
\\
\left[\frac{119}{(7h_0)100\pi L^5T\rho_\CFT^2e^{\phi_0}}\right]^{2/5} &\text{if $\mathcal{ \hat T} \ll 1$.}
\end{cases} \label{conductivityeqns}
\eea
Here we learn that the scaling of the conductivity with system size is very different in both the large and small tension cases.  As we can see, in both cases the conductivity has the same temperature scaling.  This can be understood from the fact that the changes in finite-size scaling enter entirely through the quantity $r_c$, which is independent of temperature.

\section{Free-energy calculations} \label{sec-fecalcs}

According to the rules of the AdS/CFT correspondence, to calculate the free energy of the phase corresponding to the CFT we must evaluate the gravity action on shell; {\em i.e.} at the appropriate solutions to the field equations.

\subsection*{The stellar phase}

Taking the trace of the Einstein equations
\begin{equation}
 \frac1{2\kappa^2} \left[ R + \tfrac12 (\partial\phi)^2
 - \frac{12}{L^2} \right] = - \frac{\mathcal T}{X}(X-1)^2+\frac12(3p-\rho_m) \,,
\end{equation}
and using the fact (see ref.~\cite{Hartnoll:2010gu}) that $\mathcal L^{f}_{\text{on-shell}} = p$, the free energy for the system described by a star evaluates to
\bea
  F_\ssS(T) &=& - T S_{\text{on-shell}} \nn\\
 &= & T \int \exd^4x \sqrt{-g} \left[
 \frac3{L^2\kappa^2} + \frac{\mathcal T (X-1)}{X}
 - \frac12 (\rho_m - p ) \right] \\
 &=& \frac{4\pi LT \beta'}{\kappa^2} \left\{ \int_{0}^{r_s}
 \frac{\exd r e^{a(r)/2}r^2}{\sqrt{1 + r^2 - {\kappa^2m}/
 {4\pi Lr}}} \right. \left[ 3 + \frac{\mathcal{\hat T} (X-1)}{X} + \frac{\kappa^2L^2}2 (p-\rho_m) \right] \nn\\
 &&\qquad\qquad\qquad\qquad\qquad\qquad\qquad
 + \left. \int_{r_s}^{r_\infty} \exd r e^{\xi(r)/2}
 r^2 \left[ 3 + \frac{\mathcal{\hat T} (X-1)}{X}
 \right] \vphantom{ \frac{ \exd r e^{a(r)/2}
 r^2}{\sqrt{1 +r^2 - {\kappa^2m}/{4\pi Lr}}}} \right\}
 \,. \nn
\eea

To evaluate the free energy, we numerically solve the field equations and then integrate these functions to some asymptotically large cutoff, $r_\infty$.  Since there is no horizon in this free energy to relate to a temperature, one would naively expect that this free energy is independent of temperature (since the temperature in the definition of free energy will cancel the temperature from integrating the time circle.) While this is true at low temperatures, the free energy is only meaningful when comparing to an alternative (in this case black hole) phase and appropriately normalizing the circumference of the $\mathbf{S}^1$'s at infinity as in \S\ref{sec-adsfinite}. Doing so causes the free energy of the stellar phase to behave as $\sim T^3$ in a similar way to pure AdS space at large $T$.

\subsection*{The black-hole phase}

To obtain the free energy of the black hole phase we again trace the field equations, but this time without the presence of matter, to give
\begin{equation}
 R + \frac12 (\partial\phi)^2 = \frac{12}{L^2} - \frac{2\kappa^2\mathcal T}{X}(X-1)^2 \,,
\end{equation}
and plugging this into the action gives the free energy
\bea
 F_\ssT(T) &=& - T S_{\text{on-shell}} \nn\\
 &=& T \int \exd^4 x \sqrt{-g} \left[ \frac3{L^2\kappa^2} + \frac{\mathcal T (X-1)}{X} \right] \label{BHfreeenergy}\\
 &=& \frac{4\pi L}{\kappa^2} \int_{r_h}^{r_\infty}
 \exd r \; e^{\xi(r)/2} r^2 \left[ 3 + \frac{
 \mathcal{\hat T}(X-1)}{X} \right] \,.\nn
\eea
We use these expressions to compute the free energy difference between the two phases in \S\ref{sec-phasediagram}.

\end{document}